\documentclass[fleqn,usenatbib]{mnras}

\usepackage[T1]{fontenc}

\DeclareRobustCommand{\VAN}[3]{#2}
\let\VANthebibliography\thebibliography
\def\thebibliography{\DeclareRobustCommand{\VAN}[3]{##3}\VANthebibliography}

\usepackage{graphicx}
\usepackage{subcaption}
\usepackage{amsmath}
\usepackage{amssymb}
\usepackage[dvipsnames]{xcolor} 
\usepackage{url}
\usepackage{enumitem} 
\usepackage{rotating}
\usepackage{multicol}
\usepackage{multirow}

\usepackage{subfiles}
\graphicspath{{Images/}{../Images/}} 

\newcommand{\subbib}{\bibliography{PhD, Additional_bibliography}} 

\usepackage[normalem]{ulem} 
\usepackage{wasysym}
\newcommand{\eg}[0]{$\textnormal{e.g. }$}

\newcommand{\tn}[1]{\textnormal{#1}}
\newcommand{\sub}[1]{_{\textnormal{#1}}}

\newcommand{\Msun}[0]{\,\textnormal{M}_{\textnormal{\astrosun}}}

\definecolor{darkgreen}{rgb}{0.0,0.65,0.0}

\newcommand{\oi}{[\ion{O}{i}]}
\newcommand{\oii}{[\ion{O}{ii}]}
\newcommand{\oiii}{[\ion{O}{iii}]}
\newcommand{\oiiiauroral}{[\ion{O}{iii}]$\lambda$4363}
\newcommand{\oiiauroral}{[\ion{O}{ii}]$\lambda$7320,30}
\newcommand{\sii}{[\ion{S}{ii}]}
\newcommand{\siii}{[\ion{S}{iii}]}
\newcommand{\siiiauroral}{[\ion{S}{iii}]$\lambda$6312}
\newcommand{\nii}{[\ion{N}{ii}]}

\newcommand{\Ha}{H$\alpha$}
\newcommand{\Hb}{H$\beta$}
\newcommand{\DXVI}{N2S2H$\alpha$\ }

\newcommand{\haew}{H$\alpha$EW} 
\newcommand{\HII}{H\textsc{ii}\ }
\newcommand{\Te}{$T\sub{e}$}
\newcommand{\logoh}{$12 + \log(\rm{O/H})$}
\newcommand{\logu}{$\log(U)$}
\newcommand{\hiidentify}{\texttt{HIIdentify}}
\newcommand{\hiiphot}{\texttt{HIIphot}}

\defcitealias{Pettini2004}{PP04}
\defcitealias{Marino2013}{M13}

\title[Optimal metallicity diagnostics for MUSE observations of low-z galaxies]{Optimal metallicity diagnostics for MUSE observations of low-$z$ galaxies}

\author[B. Easeman et al.]{
Bethan Easeman,$^{1}$\thanks{E-mail: be329@bath.ac.uk}
Patricia Schady,$^{1}$
Stijn Wuyts$^{1}$
and Robert M. Yates$^{2,3}$
\\
$^{1}$Department of Physics, University of Bath, Bath, BA2 7AY, United Kingdom\\
$^{2}$Centre for Astrophysics Research, University of Hertfordshire, Hatfield, AL10 9AB, UK\\
$^{3}$Astrophysics Research Group, University of Surrey, Stag Hill, Guildford, GU2 7XH, UK
}

\date{Accepted XXX. Received YYY; in original form ZZZ}

\pubyear{2023}

\begin{document}
\label{firstpage}
\pagerange{\pageref{firstpage}--\pageref{lastpage}}
\maketitle

\renewcommand{\subbib}{}

\begin{abstract}

The relatively red wavelength range (4800--9300\AA) of the VLT Multi Unit Spectroscopic Explorer (MUSE) limits which metallicity diagnostics can be used; in particular excluding those requiring the \oii$\lambda\lambda$3726,29 doublet. We assess various strong line diagnostics by comparing to sulphur \Te-based metallicity measurements for a sample of 671 \HII regions from 36 nearby galaxies from the MUSE Atlas of Disks (MAD) survey. We find that the O3N2 and N2 diagnostics return a narrower range of metallicities which lie up to $\sim$0.3~dex below \Te-based measurements, with a clear dependence on both metallicity and ionisation parameter. The \DXVI diagnostic shows a near-linear relation with the \Te-based metallicities, although with a systematic downward offset of $\sim$0.2 dex, but no clear dependence on ionisation parameter. These results imply that the \DXVI diagnostic produces the most reliable results when studying the distribution of metals within galaxies with MUSE. On sub-\HII region scales, the O3N2 and N2 diagnostics measure metallicity decreasing towards the centres of \HII regions, contrary to expectations. The S-calibration and \DXVI diagnostics show no evidence of this, and show a positive relationship between ionisation parameter and metallicity at \logoh$>8.4$, implying the relationship between ionisation parameter and metallicity differs on local and global scales. We also present \texttt{HIIdentify}, a python tool developed to identify \HII regions within galaxies from \Ha{} emission maps. All segmentation maps and measured emission line strengths for the 4408 \HII regions identified within the MAD sample are available to download.

\end{abstract}

\begin{keywords}
ISM: abundances -- HII regions -- galaxies: abundances
\end{keywords}



\section{Introduction}
\label{sect:Intro}

Gas-phase metallicity is a key indicator of how star formation has progressed through a galaxy’s history, as metals are formed and later expelled into the interstellar medium (ISM) by stars over cosmic time. A principal pursuit in galaxy evolution studies is therefore to trace the build up of heavy elements through cosmic time, and within galaxies.

The oxygen abundance, \logoh, which is typically used as a proxy for the gas-phase metallicity of the ISM, is measured via diagnostics which rely on the relative strengths of emission lines (see \eg{}\citealt{Maiolino2019}). Metal recombination lines provide the most direct measure of metallicity, and have the advantage that their line strength is weakly dependent on gas properties such as temperature and density \citep[e.g.][]{Peimbert2017}. However, the lines are very weak, being around $10^3-10^4$ times fainter than the Balmer lines \citep{Maiolino2019}, and can therefore only be detected in the very nearby Universe. Alternatively, the slightly stronger auroral lines also provide a relatively direct tracer of metallicity, whereby the relative strength of auroral to nebular lines originating from the same species provide a measure of the electron temperature (\Te) of the gas. Due to the increased cooling effect provided by the metal lines in more metal-rich gas, the metallicity and electron temperature are linked, so from this measure of \Te, it is possible to fairly accurately determine the metallicity of the gas \citep[e.g.][]{Izotov2006, Peimbert2017, Kewley2019, Yates2020}. Nevertheless, while stronger than the metal recombination lines, auroral lines are still $10-100$ times fainter than Hydrogen Balmer lines, becoming increasingly faint in more metal rich systems. This means they are also only detectable in a limited number of systems. While auroral line diagnostics can be considered a direct measurement of the physical conditions within the gas, it must be noted that they rely on a number of assumptions and simplifications \citep[e.g.][]{Perez-Montero2017, Cameron2020, Yates2020}. For example, the gas temperature is assumed to remain constant within a series of concentric shells, rather than taking into account variations on smaller scales \citep[]{Osterbrock2006, Bresolin2008}, which may lead to diagnostics under-estimating the metallicity.

Due to the faintness of these lines, strong line diagnostics have been developed, offering an essential tool to explore the gas-phase metallicity in galaxies too metal-rich or too distant for the recombination and auroral lines to be detected. These diagnostics are developed either by finding empirical relations between a combination of strong line ratios and \Te-based metallicity in \HII regions or galaxies, or equivalently, between metallicities and strong line ratios predicted with photoionisation models.

These strong line diagnostics, first developed by \cite{Alloin1979} and \cite{Pagel1979}, frequently rely on the \oii\ and \oiii\ nebular lines, such as the R23 (log((\oii$\lambda$3727 + \oiii$\lambda\lambda$4959,5007)/\Hb)) diagnostic, which uses the two principal oxygen states to account for the ionisation structure in the \HII region. This diagnostic, however, has a large dependence on the ionisation parameter, as well as being double-branched, requiring a second, less sensitive metallicity diagnostic to determine which branch applies \citep{Kewley2002,Kobulnicky2004, Maiolino2019}. The N2O2 diagnostic (log(\nii$\lambda$6584/\oii$\lambda$3727)) has very little dependence on the ionisation parameter, but primarily traces N/O, and is therefore sensitive to the assumed relation between N/O and O/H \citep{Maiolino2019}. The O3N2 (log((\oiii$\lambda$5007/\Hb)/(\nii$\lambda$6584/\Ha))) diagnostic is popular due to all relevant lines being generally accessible in a single grating setting, and its small dependence on dust reddening due to the proximity of the lines in the ratios. However, O3N2 is primarily a tracer of the ionisation parameter, \logu, thus requiring an understanding of how metallicity and ionisation parameter are related, which can vary on spatial scales and with redshift. An alternative to the O3N2 diagnostic which is similarly insensitive to dust reddening, but apparently additionally independent of \logu\ and gas pressure, is the \cite{Dopita2016} \DXVI diagnostic, which the authors thus claim is a useful diagnostic to use on high-redshift galaxies.

Thus, while strong line diagnostics are an important tool, it is important to remain mindful of their associated shortfalls, largely related to their dependency on properties other than metallicity, such as ionisation parameter, ISM pressure, and electron density \citep{Kewley2002, Dopita2016}. These limitations are manifested in the large discrepancies in metallicity that can be observed between different diagnostics \citep[e.g.][]{Kewley2008} and are not necessarily systematic. Such relative discrepancies can be seen, for example, in the shape of observed radial metallicity profiles \citep[e.g.][]{Belfiore2017,Belfiore2019, Schaefer2019, Boardman2020, Mingozzi2020, Poetrodjojo2021,Yates2021}. These discrepancies are not yet well understood \citep{Kewley2002, Stasinska2019}.

To this end, in \cite{Easeman2022}, we investigated the radial metallicity profiles of galaxies, and found large differences in the prevalence of certain features such as central dips in the radial profiles when different diagnostics were used. The prevalence of these dips in metallicity profiles measured using the O3N2 diagnostic also appeared linked to global properties of the galaxy such as stellar mass and star formation rate, whereas links to global properties were much weaker when the \cite{Dopita2016} \DXVI diagnostic was used.

Possible discrepancies between different strong line diagnostics may arise from biases present in the calibration samples used when deriving these diagnostics \citep[e.g.][]{Curti2017, Kewley2019, Stasinska2019}. Often these biases are unavoidable, for example in the need for auroral lines detections in empirically derived diagnostics, which are primarily detected in low metallicity, high excitation gas \citep{Hoyos2006}. Using diagnostics on observations of different spatial scales to the calibration sample can also be problematic. \cite{Yates2020} found that diagnostics calibrated on \HII region scales agreed well with \Te-based measurements for observations on similar scales, but less well for observations on global scales. Similarly, diagnostics calibrated on galaxy-integrated observations appeared less reliable for observations on \HII region scales. 

With higher resolution observations using integral-field unit (IFU) spectrographs such as the Multi Unit Spectrographic Explorer \citep[MUSE; ][]{Bacon2010} and, more recently, the NIRSpec integral field spectrograph on the James Webb Space Telescope \citep[JWST; ][]{JWST}, the variation in metallicity within galaxies can be studied on much smaller scales than was previously possible. The combination of high spatial resolution and large field of view has made MUSE an especially powerful instrument for studying the ISM conditions within star forming regions, and variations across a galaxy \citep[e.g.][]{Emsellem2022}. However, a drawback of MUSE is its relatively short wavelength range \citep[4650–9300 \AA;][]{Bacon2010}, which means that certain key emission lines, such as \oii$\lambda$3727,29, are not visible for low-\textit{z} galaxies, limiting the metallicity diagnostics which can be applied.

In this paper we therefore investigate the reliability of a number of strong line diagnostics when applied to MUSE data, using an empirical approach. In Section \ref{sect:Data} we detail the observations used, and present the steps taken in our analysis in Section \ref{sect:analysis}. The metallicity diagnostics used are described in Section \ref{sect:method}, and our results in Section \ref{sect:Results}, with a discussion in Section \ref{sect:discussion}. Finally, our conclusions are presented in Section \ref{sect:conclusion}.
The Appendices contain further information about the sample, as well as flux measurements for the relevant emission lines from 4408 \HII regions we identify within our sample of 36 galaxies using \hiidentify, our newly-developed python tool.

\subbib

\section{Data}
\label{sect:Data}

For our analysis we use MUSE data taken as part of the MUSE Atlas of Disks (MAD)\footnote{\url{https://www.mad.astro.ethz.ch}} survey, which covers 38 galaxies on the star forming main sequence, selected as a sample of nearby (z $<$ 0.013), relatively face-on (inclination~$<$~70$^\circ$) galaxies, with a range of masses from 10$^{8.5}$ to 10$^{11.2} \Msun$ \citep{Erroz-Ferrer2019}. 

The typical spatial resolution is $\sim$100 pc \citep{Erroz-Ferrer2019}, allowing for the study of individual \HII regions. This sample therefore provides a large number of individual \HII regions which can be identified and used in our analysis. The range of masses allows us to probe a range of metallicities, as both global and local gas-phase metallicity have been shown to correlate with stellar mass \citep{Tremonti2004,Sanchez2013}. We give details on the MAD galaxy sample in table~\ref{tab:MADprops} of the appendix.

\subbib

\section{Analysis}
\label{sect:analysis}

The IFU data returned by MUSE are 3D cubes, with 2 spatial dimensions, and one spectral dimension, meaning each pixel of the image has an associated spectrum. A number of data products have been made available from the MAD survey\footnote{\url{https://www.mad.astro.ethz.ch/data-products}}, including 2D maps of dust-corrected emission line fluxes for all strong lines within the observed wavelength range. However, for our analysis we require flux maps for additional weak, auroral emission lines, as well as associated line flux uncertainties, which are not readily available. We therefore produce our own emission line maps from the reduced MUSE data cubes, which we download from the ESO archive science portal\footnote{\url{http://archive.eso.org/scienceportal/home}}.

In order to measure accurate emission line fluxes, we first need to separate the stellar and gas emission components in order to correct for Balmer absorption from old stellar populations. A failure to correct for such stellar absorption features can result in the Balmer line fluxes being underestimated, increasingly so for bluer Balmer lines, thus affecting the measured Balmer decrement which we need to produce galaxy dust reddening maps. We use the {\sc starlight} software package to separate the stellar and gas emission components \citep{CidFernandes2005,CidFernandes2009}, following a similar procedure as described in \cite{Kruhler2017}. In summary, we use {\sc starlight} to fit a linear superposition of template spectra to each $2\times 2$ binned MUSE spectrum, using the stellar population models from \citet{Bruzual2003}. At the typical redshift of our galaxy sample ($z\sim 0.005$), the $2\times 2$ spatial binning corresponds to a physical size of $\sim$40~pc, reaching up to 100~pc for the most distant galaxy in our sample (NGC3393).
We then linearly scale the best-fit stellar template to the intensity of each of the four spaxels in our bin, and subtract this weighted stellar component from the original data to produce a gas-phase only cube. 

We removed NGC3521 from the sample, as it has very weak emission lines, and low S/N of the \oiii\ and auroral lines. NGC4593 was also removed, due to the large contribution of and active galactic nucleus in the centre, leaving us with a final sample of 36 galaxies.

\subsection{\HII region identification with \texttt{HIIdentify}}
\label{ssect:hiidentity}

\begin{figure*}
    \centering
    \begin{turn}{90}
        \centering
        \includegraphics[width=1.2\linewidth]{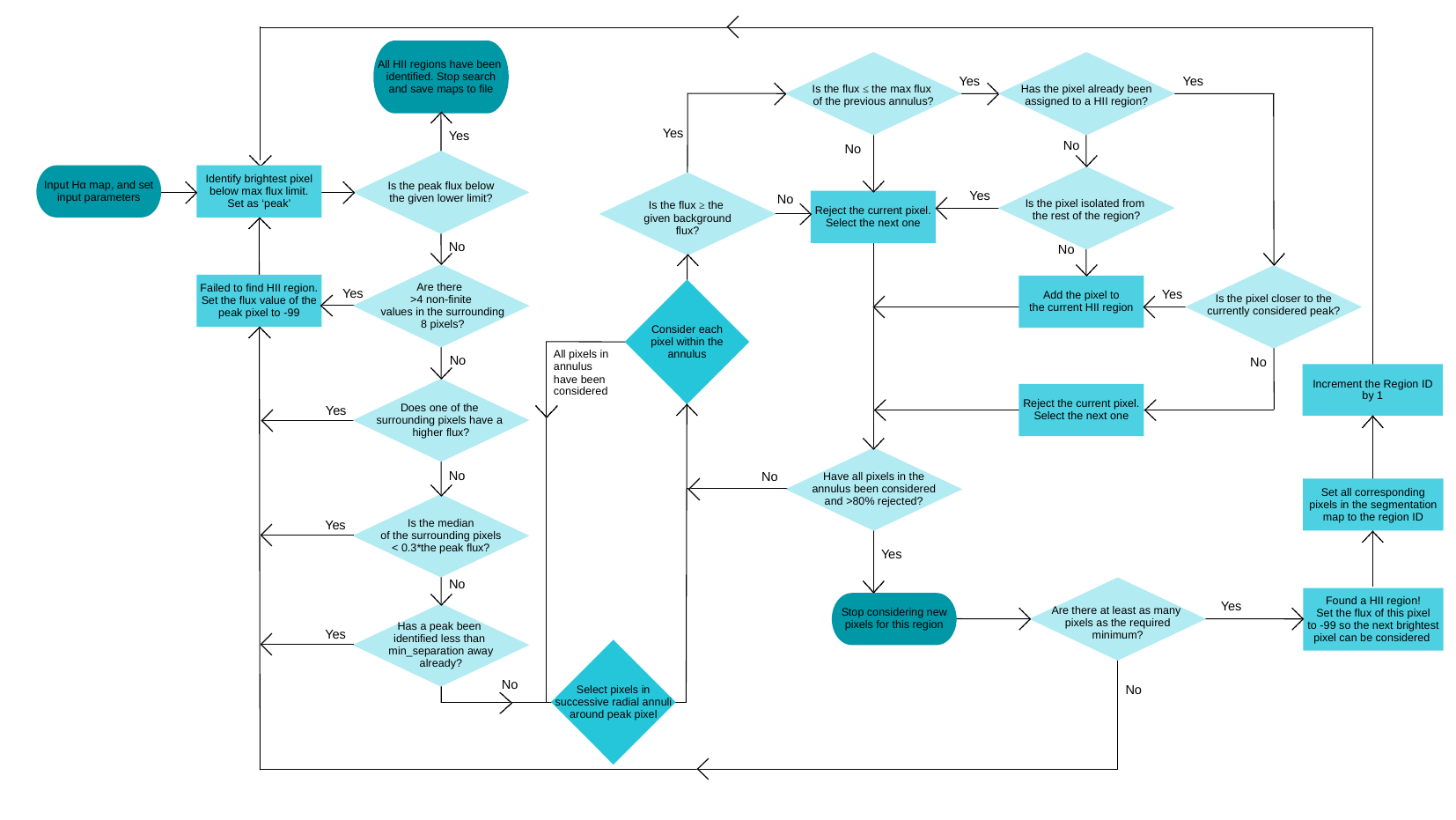}
    \end{turn}
    \caption{Flowchart illustrating the steps taken by \hiidentify{} to identify pixels as being peaks of \HII regions, removing any spaxels determined to be noise within the image, and to then assign pixels to the regions, returning segmentation maps as shown in Fig.~\ref{fig:Hamapseg}.}
    \label{fig:flowchart} 
\end{figure*}

\begin{figure*}
    \centering
    \includegraphics[width=.9\linewidth]{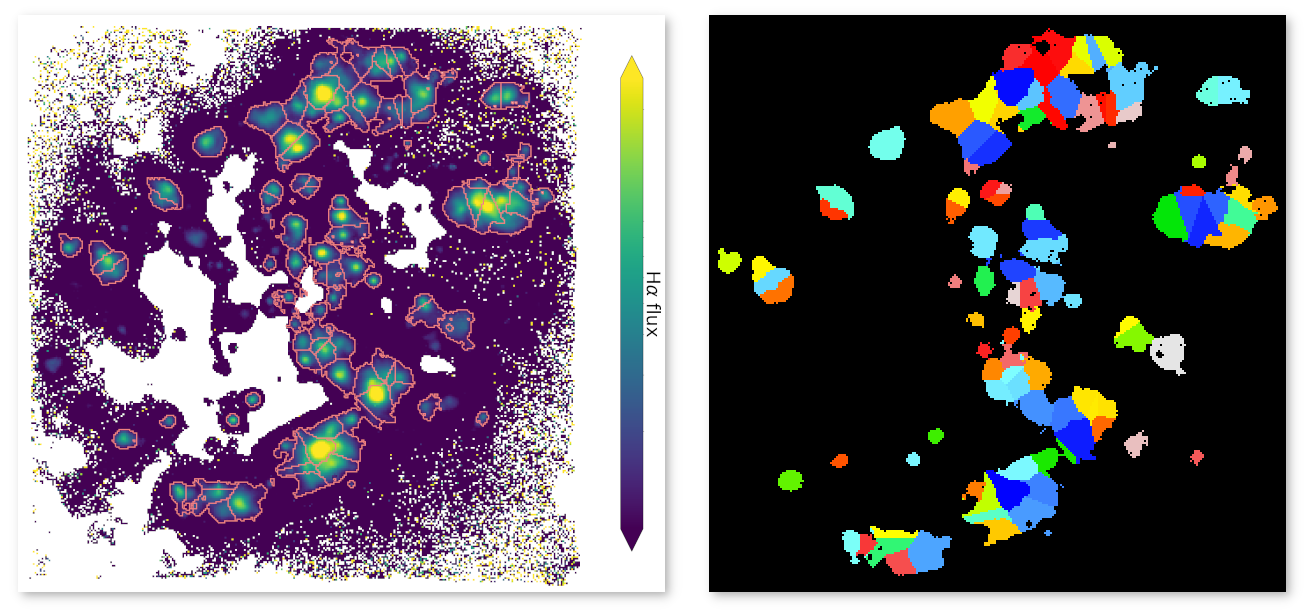}
    \caption{\textit{Left}: \Ha\ flux map of NGC1483, with outlines of the regions identified with \hiidentify{} shown in red. \textit{Right}: Segmentation map, where the pixels within each region are set to the region ID number, shown here by different colours.}
    \label{fig:Hamapseg}
\end{figure*}

To identify \HII regions within each galaxy, we use maps of the \Ha\ flux, masking out all spaxels with equivalent width of \haew $<$ 6 \AA, which are associated with Diffuse Ionised Gas (DIGs) rather than star forming (SF) regions.  DIGs have different physical properties to \HII regions, with lower gas densities, and lower ionisation parameters. The ionising source for DIGs has not been conclusively determined, meaning the metallicity diagnostics which have been calibrated against observations and models of the gas within \HII regions are not expected to remain valid when used on DIGs \citep[e.g.][]{Sanders2017, Zhang2017, Lacerda2018a}. Our choice of a threshold of 6 \AA\ is a compromise between the recommendations of \citet{Sanchez2015}, \citet{Belfiore2016} and \citet{Lacerda2018a} \citep[see also][]{Belfiore2017}. However our results are not very sensitive to the choice of the \haew\ threshold that we use, and increasing this cut to 14 \AA, as suggested by \cite{Lacerda2018a} to identify purely star-forming regions, was found to have little effect on the \HII regions identified using the process described below.

We developed a python tool, which we have named \hiidentify, to automatically identify \HII regions within a galaxy based on the \Ha{} emission, following a similar methodology to codes such as \texttt{HIIphot} \citep{Thilker2000} and \texttt{pyHIIexplorer} \citep{Espinosa-Ponce2020}. To identify \HII regions, \hiidentify{} iterates through the pixels from the brightest to the dimmest, terminating at a user-defined lower limit on the flux. Each of these pixels are considered as a possible peak of a \HII region, using several criteria to exclude noise within the image, as shown in Fig.~\ref{fig:flowchart}. For example the peak is rejected if $> 50\%$ of the surrounding pixels have non-finite flux values, or if the median flux of the surrounding pixels is $< 30\%$ of that of the peak. Other criteria must also be met for the pixel to be confirmed as the peak of a \HII region -- all of the immediately surrounding pixels must have lower fluxes than the currently considered peak, and a minimum required separation between regions can be specified. For this analysis, we used a minimum separation of 50 pc. 

Once a pixel is successfully identified as the peak of a \HII region, surrounding pixels are considered in circular annuli, and are added to the region if the flux is greater than the user-defined background flux, and the pixel has not already been added to another region. If a pixel is selected to belong to multiple regions, it is assigned to the region with the closest peak. The radius of the regions is not constrained, and instead the growth of the region stops when $> 80\%$ of the pixels in the annulus have been rejected. 

Finally, a minimum required number of pixels can be specified, which sets a lower limit on the number of pixels any identified \HII region must have. Once all possible \HII regions have been identified according to the criteria described above, our code creates a number of maps, including a segmentation map, which depicts which region each pixel belongs to, if any. \hiidentify\ is publicly available for download, with information on how to install and use it provided in the online documentation\footnote{\url{https://hiidentify.readthedocs.io/en/latest/}}. An example MUSE \Ha\ map from the MAD sample can be seen in Fig \ref{fig:Hamapseg}, with the outlines of \hiidentify{} identified regions overlaid. The associated segmentation map is also shown, indicating the spaxels belonging to each of the identified \HII regions. As there are no constraints on the shape or size of the identified regions, it can be seen that \hiidentify{} identifies all regions with peak flux above a user-defined level, and encapsulates all of the surrounding region out to a given background level, rather than making assumptions about the geometry of the regions.

The results from applying \hiidentify{} to our sample are shown in Table \ref{tab:hii_regions}, and the returned segmentation maps have been made publicly available\footnote{\url{https://doi.org/10.6084/m9.figshare.22041263}}. Input parameters were set so as to ensure that the entire \HII region was encapsulated, with any low S/N spaxels removed at a later stage. The background flux was determined using spaxels with S/N > 3, and \haew < 14 \AA, selecting the 75th percentile of the flux to represent the background level at which to set the edge of the \HII regions. Using the above input parameters, a total of 4408 \HII regions were identified in our sample of 36 galaxies, and they generally had a radius of a few hundred parsecs, which is consistent with observed sizes of \HII regions.

Following the release of the MUSE-PHANGS \citep[Physics at High Angular resolution in
Nearby GalaxieS; ][]{Emsellem2022} catalogue in \cite{Groves2023}, which included segmentation maps from the \hiiphot{} code, in Appendix \ref{app:hiidentifyhiiphot} we compare the results from the two codes, finding very good agreement in the resulting metallicity measurements.

\renewcommand{\tabcolsep}{4pt}
\begin{table}
\caption{The stellar mass ($M_{*}$), star formation rate (SFR), and number of \HII regions identified by \hiidentify{} in the 36 MAD galaxies in our sample. We also list the number of \HII regions in each galaxy that have \siiiauroral\ S/N $>$ 5 and nebular S/N > 3, and thus were included in our final sample.}
\centering
\begin{tabular}{lrrrrr}
Galaxy & \multicolumn{1}{c}{log$(M_{*})$} & \multicolumn{1}{c}{SFR} & \multicolumn{1}{c}{Num. \HII} & \multicolumn{2}{c}{S/N sel.} \\
 & \multicolumn{1}{c}{[M$_\odot$]} & \multicolumn{1}{c}{[M$_\odot$/yr]} & \multicolumn{1}{c}{regions} & \multicolumn{2}{c}{regions} \\
\hline

NGC4030    & 11.18 & 11.08 & 418 & 43 & (10\%) \\
NGC3256    & 11.14 & 3.10  & 132 & 39 & (30\%) \\
NGC4603    & 11.10 & 0.65  & 248 & 0  & (0\%)  \\
NGC3393    & 11.09 & 7.06  & 17  & 0  & (0\%)  \\
NGC1097    & 11.07 & 4.66  & 58  & 8  & (14\%) \\
NGC289     & 11.00 & 3.58  & 133 & 14 & (11\%) \\
IC2560     & 10.89 & 3.76  & 167 & 30 & (18\%) \\
NGC5643    & 10.84 & 1.46  & 95  & 10 & (11\%) \\
NGC3081    & 10.83 & 1.47  & 48  & 0  & (0\%)  \\
NGC4941    & 10.80 & 3.01  & 18  & 2  & (11\%) \\
NGC5806    & 10.70 & 3.61  & 141 & 16 & (11\%) \\
NGC3783    & 10.61 & 6.93  & 102 & 35 & (34\%) \\
NGC5334    & 10.55 & 2.45  & 122 & 12 & (10\%) \\
NGC7162    & 10.42 & 1.73  & 62  & 0  & (0\%)  \\
NGC1084    & 10.40 & 3.69  & 201 & 38 & (19\%) \\
NGC1309    & 10.37 & 2.41  & 257 & 48 & (19\%) \\
NGC5584    & 10.34 & 1.29  & 78  & 13 & (17\%) \\
NGC4900    & 10.24 & 1.00  & 352 & 43 & (12\%) \\
NGC7496    & 10.19 & 1.80  & 140 & 16 & (11\%) \\
NGC7552    & 10.19 & 0.59  & 139 & 20 & (14\%) \\
NGC1512    & 10.18 & 1.67  & 30  & 1  & (3\%)  \\
NGC7421    & 10.09 & 2.03  & 70  & 9  & (13\%) \\
ESO498-G5  & 10.02 & 0.56  & 16  & 0  & (0\%)  \\
NGC1042    & 9.83  & 2.41  & 44  & 5  & (11\%) \\
IC5273     & 9.82  & 0.83  & 94  & 17 & (18\%) \\
NGC1483    & 9.81  & 0.43  & 97  & 34 & (35\%) \\
NGC2835    & 9.80  & 0.38  & 100 & 14 & (14\%) \\
PGC3853    & 9.78  & 0.35  & 47  & 6  & (13\%) \\
NGC337     & 9.77  & 0.57  & 119 & 1  & (1\%)  \\
NGC4592    & 9.68  & 0.31  & 325 & 79 & (24\%) \\
NGC4790    & 9.60  & 0.39  & 208 & 38 & (18\%) \\
NGC3513    & 9.37  & 0.21  & 108 & 17 & (16\%) \\
NGC2104    & 9.21  & 0.24  & 90  & 11 & (12\%) \\
NGC4980    & 9.00  & 0.18  & 89  & 34 & (38\%) \\
NGC4517A   & 8.50  & 0.10  & 19  & 10 & (53\%) \\
ESO499-G37 & 8.47  & 0.14  & 24  & 8  & (33\%)\\
\hline

\end{tabular}
\label{tab:hii_regions}
\end{table}

\subsection{Spectral line fits}
\label{subsect:fitting}
To measure the line fluxes within each spaxel, we fitted the emission lines of interest with Gaussian functions using the \texttt{specutils} package in python. 

The lines were grouped into chunks with nearby lines (e.g. the \Ha\ and \nii$\lambda \lambda$6549,84 lines) when fitting, so that the positions could be tied to that of the brightest line to help constrain the fits, and in the case of doublets, the widths could be tied and the ratio of the amplitudes fixed to known theoretical ratios when fitting. The continuum extending $\sim$ 50\AA\ either side of the lines was included, allowing the continuum level to be fitted when determining line fluxes. The \oi$\lambda \lambda$6302,65 sky lines were masked out before fitting the \siiiauroral{} lines. The cube slice was dust-corrected using dust reddening maps that were produced from the measured \Ha-to-\Hb\ Balmer decrement, and assuming a \cite{Cardelli1989} attenuation law. We assumed an intrinsic Balmer decrement of 2.87 \citep{Osterbrock2006}, suitable for star forming galaxies with electron density $\sim$100cm$^{-3}$ and temperature $\sim$10,000K.

For the \siii$\lambda$9531 line, required for our \Te-based measurements, we use the known theoretical flux ratio between \siii$\lambda$9070 and \siii$\lambda$9531 of 2.47 \citep{Luridiana2015}, along with our measured flux for the \siii$\lambda$9070 line.

\subbib

\section{Metallicity and Ionisation Parameter Diagnostics}
\label{sect:method}

\Te-based methods rely on the detection of auroral lines, which are very faint compared to the strength of nebular lines - for example, the \oiiiauroral\ line is around $\sim$ 100 times fainter than the \oiii$\lambda$5007 line \citep{Maiolino2019}. Despite this difficulty, in regions where the auroral lines can be detected, \Te-based diagnostics give a more reliable measure of metallicity. We therefore focus our analysis on the subset of \HII regions where we can derive a \Te-based metallicity using the \siiiauroral{} auroral line (see Section~\ref{subsec:ZTe}), and use this as our reference metallicity to which we compare the results from a number of strong line diagnostics. Of the full sample of 4408 \HII regions, 671 had S/N > 5 detections of \siiiauroral{}, as well as S/N > 3 in all nebular lines used in the metallicity diagnostics considered in this paper. 

\subsection{\protect\textbf{\textit{T}}$\sub{\textbf{e}}$-based metallicity diagnostics} 
\label{subsec:ZTe}

As the spectral range of MUSE has a lower limit of 4650~\AA, the \oiiiauroral\ line required for calculating the electron temperature of the O$^{++}$ gas (\Te$_{,\oiii}$) is not visible for the redshift range of our sample, nor are the \oii$\lambda$3727,29 nebular lines for calculating \Te$_{,\oii}$. We are therefore unable to use a \Te-based diagnostic based on the oxygen lines. 

Sulphur has been proposed as a useful alternative tracer \citep[e.g.][]{Berg2015, Berg2020, Diaz2022}, as both sulphur and oxygen are produced in massive stars, and the yield of the two elements are expected to be linked, though with a slight time difference in their ejection from different types of supernovae \citep{Kobayashi+20}. The required \siii$\lambda$6312 auroral and \sii$\lambda$6717,31 and \siii$\lambda\lambda$9070,9531 nebular lines also experience less dust reddening due to the longer wavelengths of these lines.

In this paper we adopted the recent sulphur-based method presented in \cite{Diaz2022}.
The $R_{S3}=I(9070\AA+9531\AA)/I(6312\AA)$ line ratio is used to determine \Te$_{, \siii}$, and due to the ionisation structure of the \HII region, where the overlap between S$^{++}$ and S$^+$ appears to cover most of the region \citep{Garnett1992, Diaz2022}, it is assumed that \Te$_{, \sii}$ $\approx$ \Te$_{, \siii}$. 

To account for the contribution of S$^{3+}$ when the required [\ion{S}{iv}]$\lambda$10540 is not visible, \cite{Diaz2022} provide a method relying on the argon lines. Neither the [\ion{S}{iv}] nor argon lines are present within the wavelength range of our data, but the abundance of S$^{3+}$ is not expected to be significant in \HII\ regions within star forming galaxies such as those present within our sample \citep{Diaz2022}.

Using the measure of \Te$_{, \siii}$, and the \sii$\lambda$6317,31 and \siii$\lambda\lambda$9070,9531 nebular lines, respective values of 12 + log(S$^{+}$/H) and 12 + log(S$^{++}$/H) are determined, and combined to give 12 + log(S/H). We then convert from the 12 + log(S/H) returned from this diagnostic, to \logoh, using a fixed log(S/O) of -1.57 \citep{Asplund2009}. There have been suggestions of the S/O ratio being dependent on metallicity \citep[e.g.][]{Dors2016, Diaz2022}, though other works have found no clear dependence \citep[e.g.][]{Berg2020}. We therefore decided to use a fixed value of -1.57 in our analysis. A discussion on the implications of this choice is provided in Section \ref{subsect:SOratio}.

\subsection{Strong line metallicity diagnostics}

The relatively high lower-wavelength cut-off in the MUSE wavelength range restricts the number of diagnostics that we can use, so it was not possible to extend this analysis to commonly used diagnostics such as R23. However, this limitation makes it all the more important that an analysis is carried out to understand the robustness of metallicity diagnostics within the wavelength range and spatial scales covered by MUSE.

In this paper we test the most commonly used strong line metallicity diagnostics which include emission lines available with MUSE for nearby galaxies. Since we are interested in leveraging the high spatial resolution of MUSE, we focus on those diagnostics which have been calibrated on samples of observed or modelled \HII regions, rather than on galaxy-integrated spectra. These are the \DXVI diagnostic from \cite{Dopita2016}, the N2 and O3N2 diagnostics based on the calibrations from \cite{Pettini2004} and \cite{Marino2013}, and finally the recent S-calibration diagnostic from \cite{Pilyugin2016}.

The \cite{Dopita2016} \DXVI diagnostic, given in Eqn. \ref{eqn:D16}, was derived using the \texttt{MAPPINGS 5.0} photoionisation model code, using line ratios which are accessible from the ground out to high-redshift within a single configuration, and which have a low dependence on dust attenuation due to the proximity of the lines.
\begin{equation}
\begin{split}
    12 + \log(\tn{O/H}) =\ &8.77 + \log(\nii\lambda6584 / \sii\lambda6717,31) \\
    &+ 0.264\log(\nii\lambda6584 / \tn{H$\alpha$})
\end{split}
\label{eqn:D16}
\end{equation}

The O3N2 and N2 diagnostics are more widely used and there are a number of calibrations available. In this paper we include the \cite{Pettini2004} \citepalias{Pettini2004} calibration of O3N2 (Eq. \ref{eqn:PP04_O3N2}) and N2 (Eq. \ref{eqn:PP04_N2}), which were calibrated on a sample of 137 nearby extragalactic \HII regions with metallicity measured using predominantly \Te-based methods, but also using detailed photoionisation models at the high metallicity end. 

\begin{align}
    &&12 + \log(\tn{O/H}) &= 8.73 - 0.32 \times \tn{O3N2}\label{eqn:PP04_O3N2}\\
    &&12 + \log(\tn{O/H}) &= 8.90 + 0.57 \times \tn{N2}
    \label{eqn:PP04_N2}
\end{align}

We also consider the more recent O3N2 and N2 re-calibrations from \cite{Marino2013} (\citetalias{Marino2013}), shown in Eqs. \ref{eqn:M13_O3N2} \& \ref{eqn:M13_N2}, which used a larger sample of 603 \HII regions from the literature, along with 16 \HII complexes from the Calar Alto Legacy Integral Field Area Survey \citep[CALIFA; ][]{Sanchez2012a} with available \Te-based metallicities.

\begin{align}
    &&12 + \log(\tn{O/H}) &= 8.533 - 0.214 \times \tn{O3N2}
    \label{eqn:M13_O3N2}\\
    &&12 + \log(\tn{O/H}) &= 8.743 + 0.462 \times \tn{N2}
    \label{eqn:M13_N2}
\end{align}

Finally, we also use the S-calibration from \cite{Pilyugin2016}, which uses a sample of 313 \HII regions with spectra from single-slit observations to derive their relationships, with measured \Te-based metallicities spanning 7.0 < \logoh\ < 8.8.  The S-calibration is derived for use when the \oii$\lambda \lambda$3727,29 lines are unavailable, as is the case for our MUSE data, and instead uses the \sii$\lambda \lambda$6717,31 lines. They find very good agreement (within $\sim$0.05 dex) between their S-calibration and the R-calibration, which uses the \oii\ lines for single-slit spectra from a test sample of over 3000 \HII regions. Of note is that \cite{Pilyugin2022} find that when comparing spectra of \HII regions from IFU to single-slit observations, the S-calibration underestimates metallicities by $\sim$0.06 dex on average at \logoh\ $\gtrsim$ 8.55, and by $\sim$ 0.02 dex on average at lower metallicities. We discuss the potential impact of this on our results in Section \ref{sect:discussion}. Both the S- and R-calibration are double-branched, and \cite{Pilyugin2016} thus recommend using the line ratio log($N_2$) (where $N_2$ = \nii$\lambda \lambda$6548,84 / \Hb) to separate the upper and lower branch. For the upper branch (log($N_2$) $\geq$ -0.6), the relationship in Eq. \ref{eqn:Scal_upper} is used; for the lower branch, the relationship in Eq. \ref{eqn:Scal_lower}. Here $S_2$ = \sii$\lambda \lambda$ 6717,31 / \Hb, and $R_3$ = \oiii$\lambda \lambda$4959,5007 / \Hb.

\begin{equation}
    \begin{split}
    12 + & \log(\tn{O/H}) = 8.424 + 0.030 \log(R_3/S_2) + 0.751 \log(N_2) + \\
    &(-0.349 + 0.182 \log(R_3/S_2) + 0.508 \log(N_2)) \times \log(S_2)
    \end{split}
    \label{eqn:Scal_upper}
\end{equation}
\begin{equation}
    \begin{split}
    12 + & \log(\tn{O/H}) = 8.072 + 0.789 \log(R_3/S_2) + 0.726 \log(N_2) + \\
    &(1.069 - 0.170 \log(R_3/S_2) + 0.022 \log(N_2)) \times \log(S_2)
    \end{split}
    \label{eqn:Scal_lower}
\end{equation}

\subsection{Ionisation parameter diagnostic}

The O3N2 and N2 diagnostics are strongly dependent on the ionisation parameter (\logu), a measure of the central ionising source's ability to ionise the surrounding gas. Some works have suggested that these diagnostics may become unreliable under certain \logu\ conditions if, for example, these conditions are not represented in the calibration samples \citep[e.g.][]{Kruhler2017, Mao2018}. Therefore, we explore whether there is any evidence of this in our data.

To measure \logu, we used the \cite{Mingozzi2020} sulphur-based diagnostic, calibrated on spatially resolved IFU data. Sulphur-based ionisation parameter diagnostics have been found to have a lower dependence on the metallicity compared to oxygen-based diagnostics \citep{Kewley2002,Dors2011}, making the sulphur-based diagnostic in principle more reliable, but also suggesting that any observed trends between metallicity and \logu{} should not be a consequence of the \logu{} diagnostic itself.

This diagnostic relies on the ratio of the \siii$\lambda\lambda$9070,9531 and \sii$\lambda\lambda$6717,31 line doublets. The \cite{Mingozzi2020} re-calibration accounts for an overestimate of the strength of the \siii\ lines, suggested by \cite{Kewley2002} to be the cause of sulphur-based diagnostics under-predicting the ionisation parameter compared to oxygen-based diagnostics.

\subbib

\section{Results}
\label{sect:Results}

To investigate the diagnostics on \HII region scales, we identified \HII regions using \hiidentify, and stacked the spectra of all selected spaxels within each region. We then identified the sub-sample of regions with S/N of the measured \siiiauroral\ auroral line > 5, to allow for a reliable measurement of the \Te-based metallicity. We imposed an additional criteria to ensure that the nebular lines needed in the strong line metallicity diagnostics that we used had a S/N > 3 within each \HII stacked spectrum. This left us with a sample of 671 \HII regions within a total of 31 galaxies, with logarithmic stellar masses ranging between $\log(\rm{M}_\star/\rm{M}_\odot)=8.5$ and $\log(\rm{M}_\star/\rm{M}_\odot)=11.2$, and SFRs in the range 0.1-11.1 M$_\odot$/yr. The galaxies NGC4603, NGC3393, NGC3081, NGC7162, and ESO498-G5 had no \HII regions which met the S/N criteria (see Table~\ref{tab:hii_regions}).

\subsection{Sulphur-based \protect\textbf{\textit{T}}$\sub{\textbf{e}}$ metallicity diagnostics}

\begin{figure*}
    \centering
    \includegraphics[width=1\linewidth]{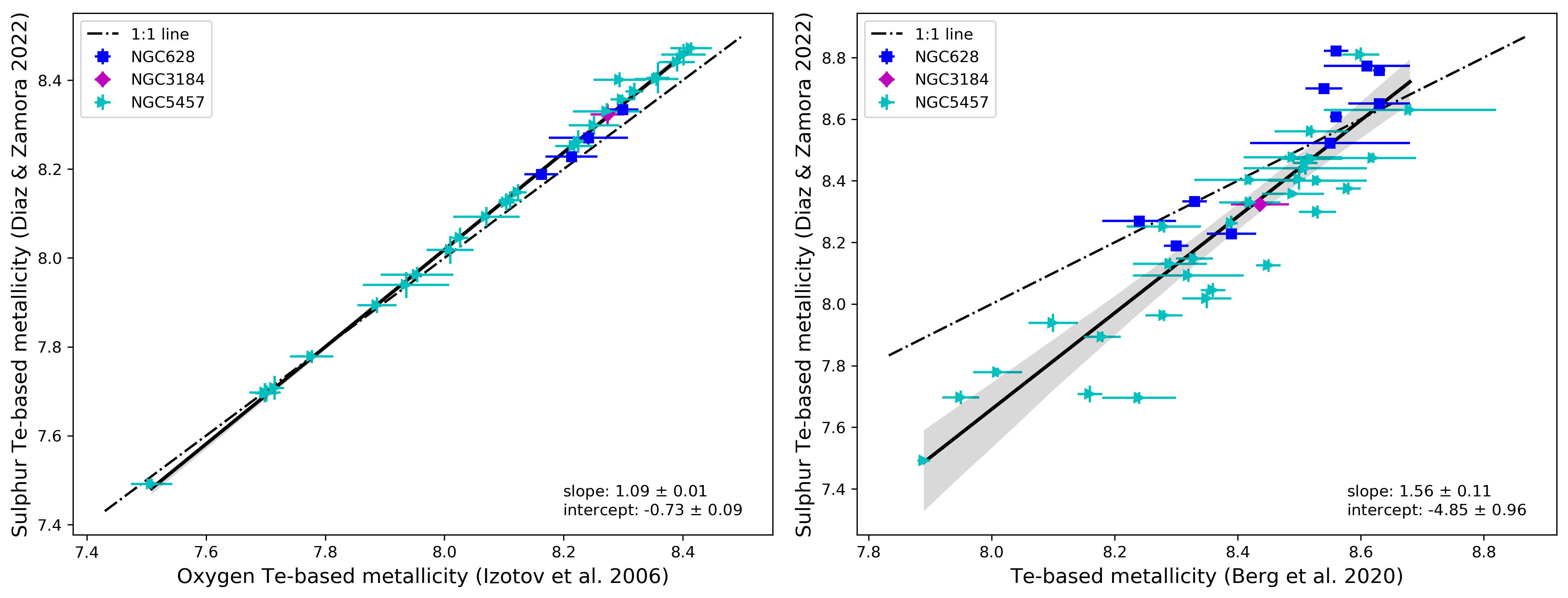}
    \caption{Comparison of the \protect\cite{Diaz2022} sulphur-based \Te\ diagnostic, to other \Te-based diagnostics. In both panels, the dot-dashed black line shows the 1:1 relationship, and the solid black line the best-fit to the data, with the shaded region representing the uncertainty on the fit. Comparison to the results from the \protect\cite{Izotov2006} diagnostic shows a good agreement (\textit{left}), but when comparing to the published \protect\cite{Berg2020} values (\textit{right}), a systematic offset at lower metallicities is seen. The slope and intercept of the best-fit line are given in the bottom right of each panel. Note that not all \HII regions within the CHAOS sample have the necessary oxygen auroral line detections required to apply the \citet{Izotov2006} diagnostic, hence the range in y-axis values differ between the two panels.}
    \label{fig:CHAOS_Te_comparison}
\end{figure*}

As all of our empirically-derived strong line metallicity diagnostics were calibrated against \Te-based diagnostics using the \oiiiauroral\ line, we first investigate the agreement between \Te-based diagnostics that rely on the \oiiiauroral\ and the \siiiauroral\ lines, before comparing our \siiiauroral\ \Te-based metallicities to the strong line diagnostics.

Since the \oiiiauroral\ line is not visible in our MUSE data, to make these comparisons we used the published data of \HII regions taken with the Multi-Object Double Spectrographs on the Large Binocular Telescope \citep[MODS;][]{Pogge10} as part of the CHemical Abundances Of Spirals (CHAOS) project \citep{Berg2015}. These observations provide a broad wavelength coverage, extending from 3200\AA\ to 10,000\AA, enabling the detection of multiple auroral lines, including \oiiiauroral, out to $z\sim 0.36$. Thus far the multi-slit data for four star-forming galaxies observed as part of CHAOS have been published, amounting to published spectra for 190 \HII regions \citep{Berg2015,Croxall15,Croxall16,Berg2020}.

For this comparison we apply the methodology described in Section \ref{subsec:ZTe} to the CHAOS data to measure the metallicities using the sulphur \Te-based diagnostic from \cite{Diaz2022}, as well as applying the \citet{Izotov2006} \oiiiauroral-based method. In addition, we compare these sulphur and oxygen-based \Te\ values to the metallicities presented in \cite{Berg2020}, who use a multiple auroral line method to measure the temperature in a 3-zone temperature \HII model that incorporates low, intermediate, and high ionisation zones. They probe the temperatures (and thus metallicities) of each of these zones using a combination of the \nii, \siii, and \oiii\ auroral lines. These metallicities are then combined to obtain the total metallicity of the region.

In order to compare the published \citet{Berg2020} \HII region metallicities to the respective \citet{Izotov2006} and \citet{Diaz2022} oxygen and sulphur \Te-based metallicities, we selected all \HII regions from the CHAOS sample with \oiiiauroral\ and \oiiauroral\ lines detected with S/N > 3 \citep[as was done by][]{Berg2020}, and with \siiiauroral\ lines detected with S/N > 5. This more stringent criteria on the S/N of the \siiiauroral\ line was used to remain compatible with the selection criteria used in our MAD sample, though we find it makes no clear systematic difference to use a cut of \siiiauroral\ S/N > 3 instead. NGC5194 had no regions with detected \oiiiauroral, leaving a total of 43 regions within the remaining three CHAOS galaxies which meet our S/N criteria.

Fig.~\ref{fig:CHAOS_Te_comparison} (left panel) shows the comparison between the sulphur-based \citep{Diaz2022} method and the oxygen \citep{Izotov2006} based method. We find very good agreement between these diagnostics, suggesting that the sulphur-based diagnostic is appropriate as a tracer of the oxygen abundance for verifying the accuracy of strong line diagnostics in our MAD sample. The line of best-fit (black line in Fig.~\ref{fig:CHAOS_Te_comparison}, left panel) and small scatter suggest a tight positive relationship between the two methods, with the sulphur-based diagnostic returning slightly higher metallicity values for \logoh{}$\gtrsim$ 8.0, leading to an offset of $\sim$0.02 dex at \logoh{}= 8.4. For the CHAOS galaxies, \cite{Berg2020} compare \Te{}[SIII] and \Te{}[OIII], and find that at low temperatures (corresponding to high metallicity), \Te{}[SIII] is shifted to lower values than \Te{}[OIII]. This may explain the offset of the sulphur-based diagnostic to slightly higher values at high metallicities.

However, when comparing the \siiiauroral\ based metallicity to the multi-zone metallicities from \cite{Berg2020}, the data no longer fall along the one-to-one line (dot-dashed line in right panel, Fig.~\ref{fig:CHAOS_Te_comparison}). There is a systematic offset at low metallicities where the \cite{Berg2020} method returns higher metallicity values, with the sulphur-based measurements offset by -0.37 dex at \logoh{}= 8.0. Nevertheless, although there is not a one-to-one agreement between the results of the two diagnostics, there is a clear relation, making it possible to convert between the results from the two diagnostics using the parameters of our best fit line, with slope = 1.56 $\pm$ 0.11, and y-intercept = -4.85 $\pm$ 0.96. The implications of applying such a re-scaling to our sulphur-based \Te{}-based metallicities are discussed in Section~\ref{subsec:SLVsTe}.

\subsection{Comparing strong line and \protect\textbf{\textit{T}}$\sub{\textbf{e}}$-based diagnostics} 
\label{subsec:SLVsTe}

\begin{figure*}
    \centering
    \includegraphics[width=1\linewidth]{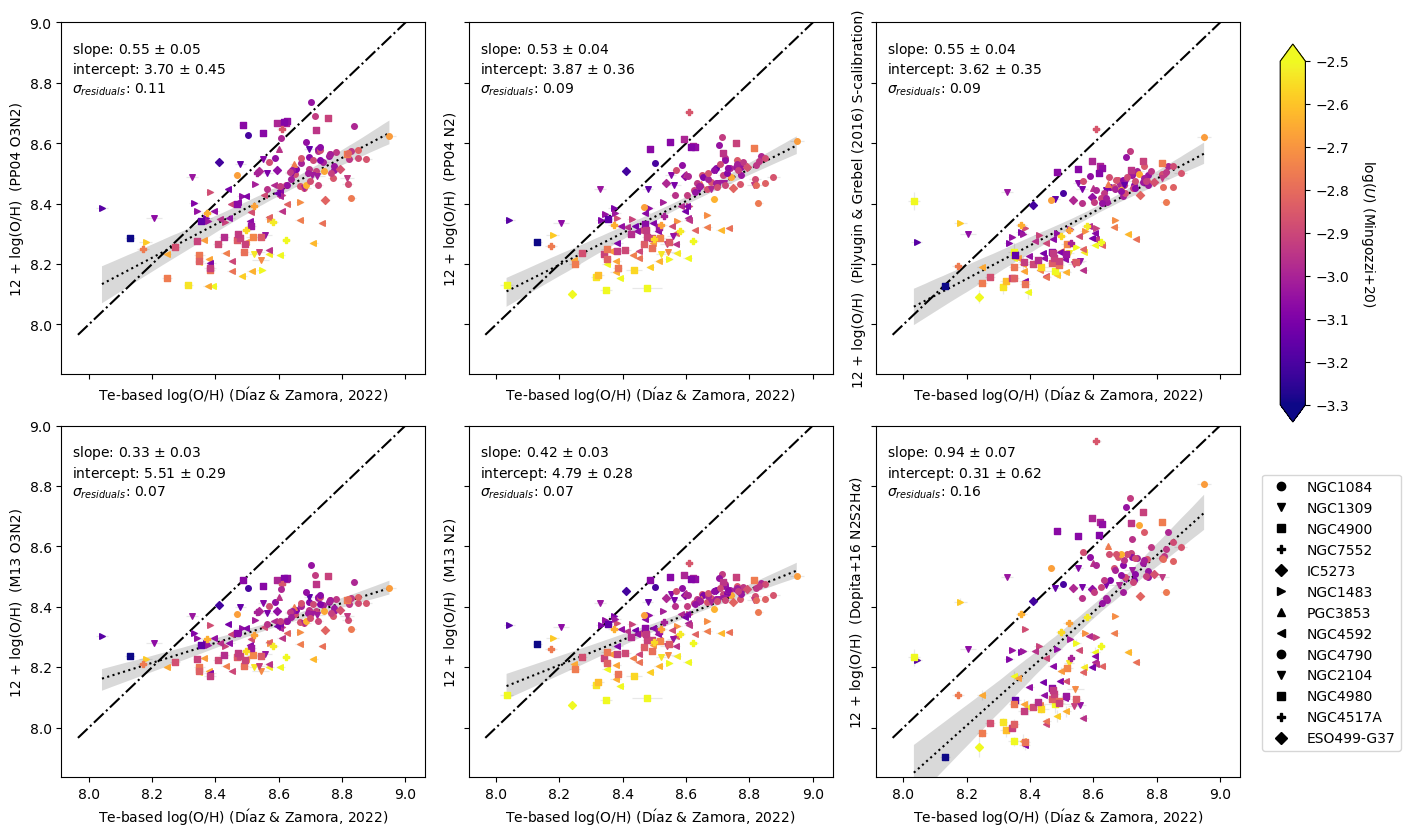}
    \caption{Comparison between strong line and sulphur \Te-based metallicity diagnostics, with the dot-dashed line showing a 1:1 relationship between the diagnostics, and a dotted line representing the best-fit. The points are coloured by \logu, measured using the sulphur-based \protect\cite{Mingozzi2020} diagnostic. The scatter about the best-fit line is denoted by the standard deviation of the residuals ($\sigma_{residuals}$).}
    \label{fig:SLvsTe}
\end{figure*}

Having found the \cite{Diaz2022} sulphur-based \Te{} method to agree well with the oxygen-based \Te{} method, we can now explore the reliability of applying strong-line diagnostics to our MUSE observations, by comparing the results from each strong-line diagnostic to that from the \cite{Diaz2022} \Te-based method. These comparisons are shown in Fig.~\ref{fig:SLvsTe}, with the points coloured by \logu. The dot-dashed line shows the 1:1 relationship between the diagnostics, and the dotted line represents the best fit, obtained using the seaborn \texttt{regplot} method, with consistent fits obtained using a bootstrap method.

According to the best-fit relationships, a number of the diagnostics show significant disagreement with the \Te-based measurements, particularly at high metallicity. The \citetalias{Pettini2004} O3N2, \citetalias{Pettini2004} N2, and \cite{Pilyugin2016} S-calibration all show some agreement with the \Te-based measurements at low metallicities, but show increasing discrepancies at higher metallicities. For these diagnostics, the best-fit line to the data returns a slope of $\sim$0.55, leading to the strong line diagnostics returning results $\sim$ 0.3 dex lower than the \Te-based measurements at \logoh\ = 8.8. One suggestion for the offsets observed with the \citetalias{Pettini2004} diagnostics could be due to the relatively small calibration sample used. The auroral line becomes increasingly faint at higher metallicities, which results in this region being less well sampled. This is evident in the \citetalias{Pettini2004} sample, where the low O3N2, high oxygen abundance region of the parameter space contains just six data points, four of which have metallicities based on strong line diagnostics. The authors themselves note the need to increase the sample size of their calibration sample at the high metallicity end.

The \citetalias{Marino2013} O3N2 and N2 diagnostics show slightly reduced scatter, but the least agreement with the \Te-based measurements, with an offset of $\sim$ 0.4 dex below the \Te-based metallicities at \logoh\ = 8.8.  The \DXVI diagnostic returns an average offset of $\sim$0.2 dex below the \cite{Diaz2022} \Te{}-based metallicities, but with a slope consistent with unity (see Table~\ref{table:bestfit}).

As expected, the highly \logu-sensitive O3N2 and N2 diagnostics show a clear variation of \logu{} along the y-axis, with the lowest metallicity regions as measured by the strong line diagnostics having higher \logu. This leads to the diagnostics showing an increased discrepancy at higher \logu. Unfortunately, the line fluxes necessary to determine \logu\ for the \citetalias{Pettini2004} sample were not published in the original paper, and we therefore cannot verify this potential \logu-bias in the \citetalias{Pettini2004} calibration sample.

Interestingly, considering the variation of \logu{} along the x-axis, we see no clear trend between \Te-based metallicity and \logu, suggesting the O3N2 and N2 diagnostics may become unreliable under certain \logu{} conditions. The \cite{Pilyugin2016} diagnostic shows a weaker trend with \logu, and the \DXVI diagnostics shows no evidence of a relationship. 

\begin{figure*}
    \centering    
    \includegraphics[width=1\linewidth]{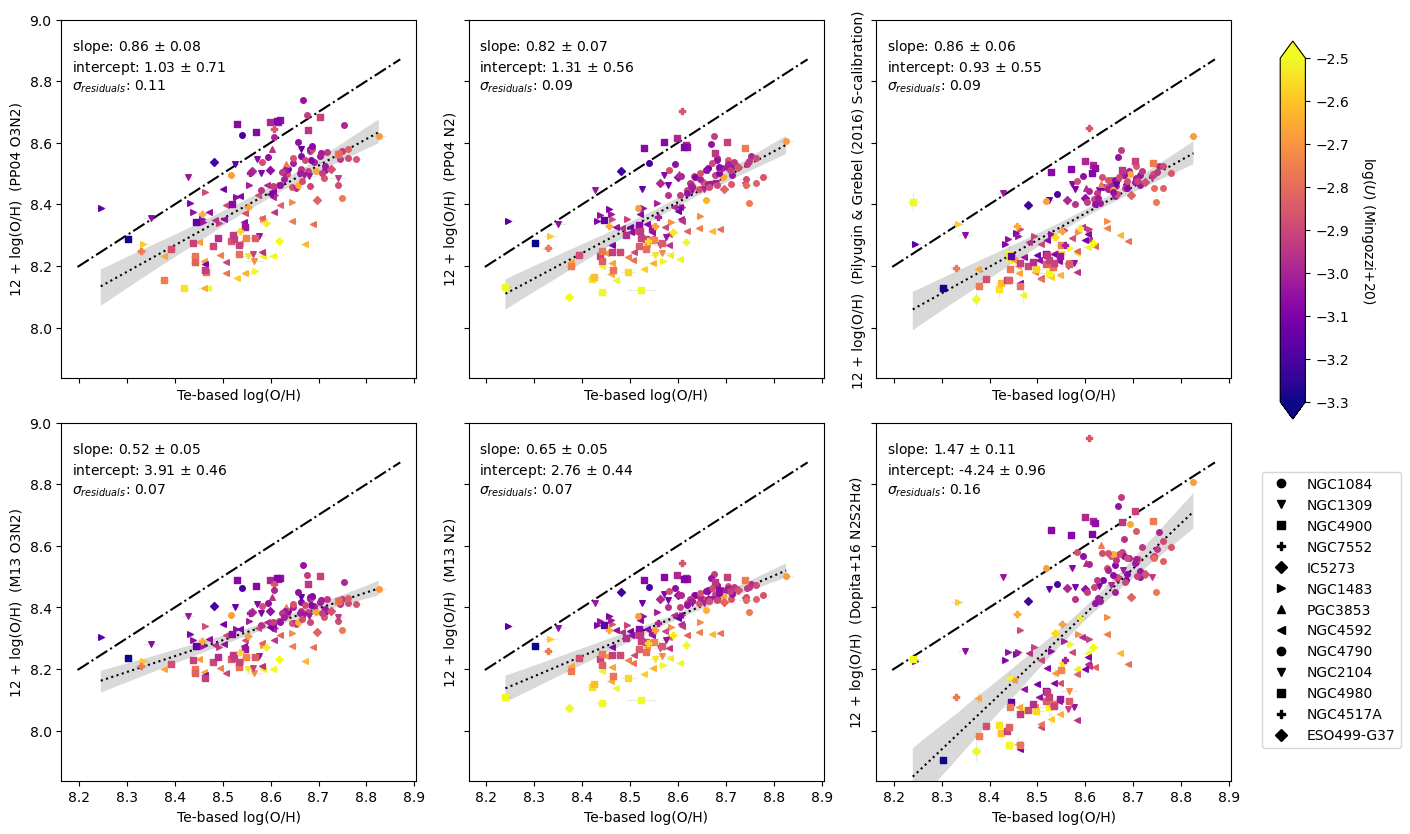}
    \caption{As for Fig.~\ref{fig:SLvsTe}, but using the relationship between the \protect\cite{Diaz2022} and \protect\cite{Berg2020} \Te-based diagnostics shown in Fig.~\ref{fig:CHAOS_Te_comparison} to convert from the measurements made using the \protect\cite{Diaz2022} diagnostic to the predicted value for if the \protect\cite{Berg2020} diagnostic could be used.}
    \label{fig:SLvsTe_Berg_conversion}
\end{figure*}

If the arguably more sophisticated multi-zone \cite{Berg2020} metallicity diagnostic is considered the more accurate of the \Te-based diagnostics investigated in this work, this would imply that our sulphur \Te-based metallicities are underestimated at low metallicities. Although this would not improve the discrepancies seen at high metallicity between our \Te-metallicities and the strong line diagnostics shown in Fig.~\ref{fig:SLvsTe},  it could, in part, explain the sub-linear relation observed in most cases. To explore this further, we use our best-fit relationship between the sulphur \Te-based metallicities and the \cite{Berg2020} metallicities shown in Fig.~\ref{fig:CHAOS_Te_comparison} to convert our sulphur-based \Te-metallicities to the \cite{Berg2020} metallicity scale, and compare this to the strong line metallicities in Fig.~\ref{fig:SLvsTe_Berg_conversion}. As expected, the \citetalias{Pettini2004} O3N2 and N2, and the S-calibration metallicities show a more linear relation with the \Te-based values (best-fit slopes increase from $\sim 0.5$ to $\sim 0.8$), but with a systematic offset to lower metallicities of $\sim$0.1 dex at \logoh=8.2, and $\sim$0.2 dex at \logoh{}= 8.8. Although the relations with the \citetalias{Marino2013} diagnostics also steepen, they still remain fairly flat, with a slope in the range $0.5-0.6$ (see Table~\ref{table:bestfit}). The \DXVI diagnostic, however, shows a much lower level of agreement at low metallicity when converting to the predicted \cite{Berg2020} values, with a best-fit slope of $\sim$1.5, and larger offsets of $\sim$0.4 dex at \logoh{}= 8.2. The best-fit parameters are summarised in Table~\ref{table:bestfit}. 

Although we cannot conclusively state which of the \Te-metallicity diagnostics is the more accurate, it is noteworthy that the \citetalias{Pettini2004} and \citetalias{Marino2013} O3N2 and N2 diagnostics, and the \cite{Pilyugin2016} diagnostic were all calibrated against two-zone temperature models. We would therefore expect these diagnostics to be in better agreement with the \cite{Diaz2022} sulphur-based \Te-metallicity, which is in far better agreeement with the \cite{Izotov2006} oxygen two-zone temperature model (see Fig.~\ref{fig:CHAOS_Te_comparison}) than the \cite{Berg2020} three-zone temperature model. Furthermore, the offsets at high metallicity apparent in Fig.~\ref{fig:SLvsTe} between \Te-metallicities and the N2, O3N2 and S-calibration can be explained to some degree by selection effects in the calibration samples. We discuss this further in section~\ref{ssubsect:TeSelectionEffects}.

\begin{table*}
\caption{Parameters for the best-fit lines shown in Figs. \ref{fig:SLvsTe} and \ref{fig:SLvsTe_Berg_conversion}, such that \logoh$_X$ = slope $\times$ \logoh$_{T\rm{e}}$ + intercept, where $X$ is the strong line diagnostic.}
\centering
\begin{tabular}{l|c|c|c|c}

Strong line & \multicolumn{2}{c}{\cite{Diaz2022} \Te-based} & \multicolumn{2}{c}{Predicted \cite{Berg2020}} \Te-based \\
diagnostic & slope & intercept & slope & intercept \\
 \hline \\

 PP04 O3N2 & 0.55 $\pm$ 0.05 & 3.70 $\pm$ 0.45 & 0.86 $\pm$ 0.08 & 1.03 $\pm$ 0.71\\

 PP04 N2 & 0.53 $\pm$ 0.04 & 3.87 $\pm$ 0.36 & 0.82 $\pm$ 0.07 & 1.31 $\pm$ 0.56 \\

 S-calibration & 0.55 $\pm$ 0.04 & 3.62 $\pm$ 0.35 & 0.86 $\pm$ 0.06 & 0.93 $\pm$ 0.55 \\

 M13 O3N2 & 0.33 $\pm$ 0.03 & 5.51 $\pm$ 0.29 & 0.52 $\pm$ 0.05 & 3.91 $\pm$ 0.46 \\

 M13 N2 & 0.42 $\pm$ 0.03 & 4.79 $\pm$ 0.28 & 0.65 $\pm$ 0.05 & 2.76 $\pm$ 0.44 \\

 \DXVI & 0.94 $\pm$ 0.07 & 0.31 $\pm$ 0.62 & 1.47 $\pm$ 0.11 & -4.24 $\pm$ 0.96 \\
 \hline

\label{table:bestfit}
\end{tabular}
\end{table*}

\subsection{Comparing to the Curti et al. O3N2 re-calibration}

As the O3N2 diagnostics presented in Fig.~\ref{fig:SLvsTe} show significant discrepancies with the results from the \Te-based diagnostic, and given the potential selection effects in the older calibrations, we investigate the impact of using the more recent \cite{Curti2017} re-calibration of the O3N2 diagnostic. Although this re-calibration was based on stacked galaxy-integrated spectra, and it is preferable to use metallicity diagnostics which have been calibrated on data that probe similar spatial scales to the science data, this re-calibration included a very large, and arguably unbiased calibration sample. Using a consistent underlying calibration sample, \cite{Curti2017} re-calibrated six diagnostics, of which the $R_3$ ([\ion{O}{iii}]$\lambda$5007/H$\beta$), $N_2$ ([\ion{N}{ii}]$\lambda$6584/H$\alpha$) and O3N2 all contain nebular emission lines probed by our MUSE data. However, we focus our analysis only on the more commonly used O3N2 diagnostic.

This re-calibration used \oiiiauroral\ line fluxes obtained from whole-galaxy observations, with spectra stacked in bins of log(\oii /\Hb) and log(\oiii /\Hb) to obtain sufficient S/N of the auroral line, in particular at higher metallicities. The sample of galaxies was selected from SDSS data release 7 \cite[DR7][]{aaa+09} and had a median redshift of 0.072, leading to each fibre corresponding to observations on the scale of $\sim$3 kpc. Using the oxygen \Te-based metallicities, \cite{Curti2017} then modelled the relation between the strong line ratios and the \Te-based metallicities with a high order polynomial. In the case of the O3N2 diagnostic, their best-fit polynomial was of the form

\begin{equation}
{\rm O3N2} = \sum_{n=0}^2 c_n x^n,
\end{equation}
where $x$ is the oxygen abundance normalised to the Solar value \citep[12+log(O/H)$_\odot$=8.69;][]{Asplund2009}, and $c_0=0.281$, $c_1=-4.765$ and $c_2=-2.268$.

\begin{figure}
    \centering\includegraphics[width=1\linewidth]{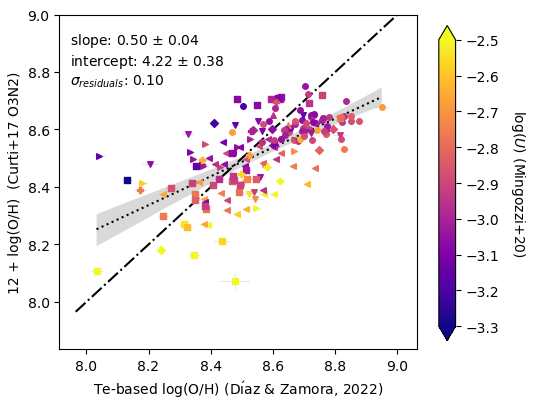}
    \caption{As for Fig.~\ref{fig:SLvsTe}, showing the comparison of the \protect\cite{Curti2017} O3N2 against the \protect\cite{Diaz2022} \Te-based diagnostic.}
    \label{fig:SLvsTe_Curti}
\end{figure}

Fig.~\ref{fig:SLvsTe_Curti} shows that the average difference between the strong line and \Te-based diagnostics is reduced when the \cite{Curti2017} O3N2 diagnostic is used, compared to the \citetalias{Pettini2004} or \citetalias{Marino2013} calibrations, because the \cite{Curti2017} O3N2 values are generally shifted to higher metallicities. However, the line of best fit still features a sub-unity slope of $\sim{}0.50$, similar to the \citetalias{Pettini2004} O3N2, N2, and the S-calibration results. This suggests that the increased discrepancy between the strong line and \Te-based diagnostics at high metallicity cannot simply be explained by biases in the calibration sample. 

The \cite{Curti2017} calibration sample, by using observations of galaxy-integrated spectra, averages over a large number of \HII regions in each observation, thus averaging over a range of different ionisation conditions. Despite this, there remains a clear \logu{} dependence in the \cite{Curti2017} metallicities, again suggesting that the \logu{} must be taken into account in order to reduce the scatter in O3N2 metallicities.

\begin{figure}
    \centering
    \includegraphics[width=1\linewidth]{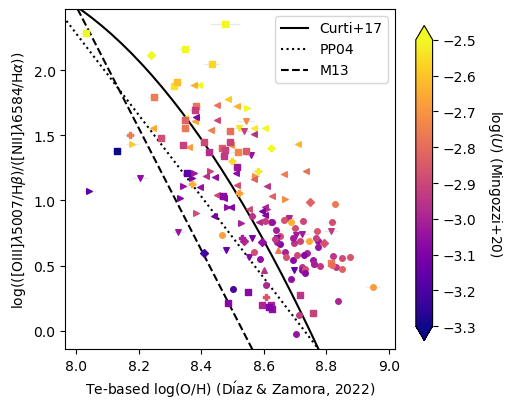}
    \caption{The O3N2 line ratio against sulphur \Te-based metallicity, colour coded by \logu. The best fit relations from \protect\citetalias{Pettini2004} (dotted), \protect\citetalias{Marino2013} (dashed), and \protect\cite{Curti2017} (solid) are overplotted. The differences between the O3N2 diagnostics can be clearly seen, with the \protect\citetalias{Pettini2004} and \protect\citetalias{Marino2013} diagnostics shifted to lower metallicities than the \protect\cite{Curti2017} re-calibration. The \protect\citetalias{Marino2013} diagnostic also clearly shows a steeper relationship, hence covering a much smaller range of metallicities.}
    \label{fig:O3N2lineratio}
\end{figure}

The relatively flat relation between \Te-based metallicities and O3N2 and N2 diagnostics implies that the strong line ratios used in these diagnostics are not sufficiently sensitive to metallicity, leading to a narrower range in metallicity estimates than \Te-based methods. This can be understood from the unaccounted \logu{} dependency in the O3N2 and N2 diagnostics \citep{Kewley2002}, which are effectively flattened out in the calibrations. To illustrate this more clearly, in Fig.~\ref{fig:O3N2lineratio} we show the O3N2 line ratio against \Te-based metallicity for our sample of \HII regions, colour-coded by \logu, and we overplot the best-fit relation between these two parameters from \citetalias{Pettini2004}, \citetalias{Marino2013} and \cite{Curti2017}. From this figure it can be clearly seen how \HII regions with comparable O3N2 line ratios can have very different metallicities, depending on their ionisation parameter, as expected from \cite{Kewley2002}. 

\begin{figure*}
    \centering
    \includegraphics[width=1\linewidth]{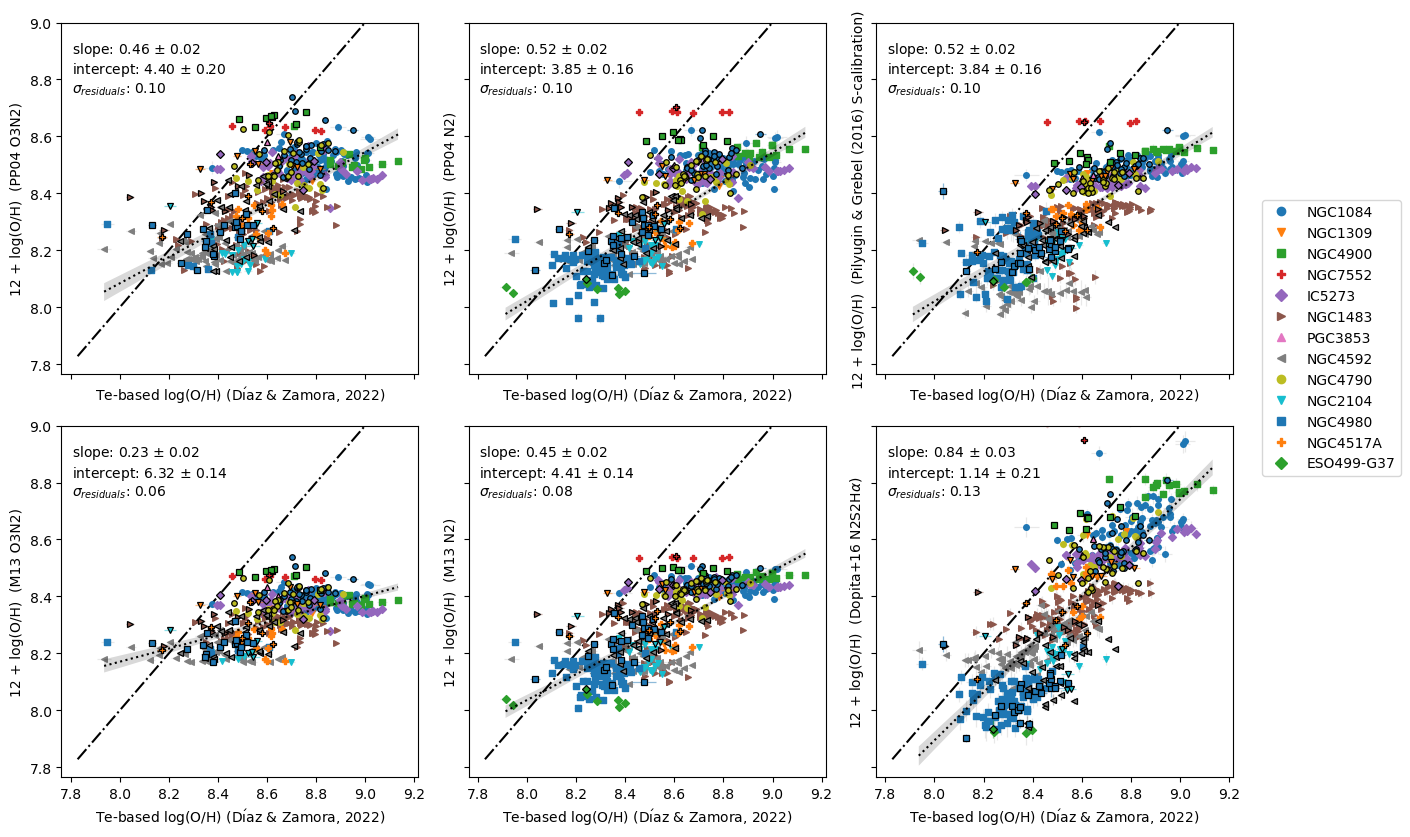}
    \caption{As for Fig.~\ref{fig:SLvsTe}, now showing the metallicities of individual spaxels (no outlines), together with the metallicities from whole \HII region stacked spectra overlaid with black outlines. The colour and shape of the markers is unique to each galaxy. The best-fit line, with the uncertainty shown by the shaded region, is fitted only to the spaxel measurements. }
    \label{fig:SLvsTe_spaxels} 
\end{figure*}

\subsection{Metallicity diagnostics on sub-\HII region scales}
\label{subsect:spaxels}

The \HII regions identified with our \hiidentify\ code range in size from 22 pixels up to 1740 pixels, corresponding to 0.08 -- 0.98 kpc. Therefore, in addition to investigating the properties of individual \HII regions, the MUSE data used in this paper also allow us to explore the range of metallicity and \logu\ values \textit{within} \HII regions. This may provide further insight into how representative H\textsc{ii}-integrated properties are of the range in values within the \HII region.

Using the same sample of \HII regions with measured \Te-based metallicities as in our previous analysis, we selected the spaxels belonging to these \HII regions, as determined using \hiidentify. We repeated our previous analysis using these spaxels, and in Fig.~\ref{fig:SLvsTe_spaxels} we compare the results from the strong line diagnostics to the \cite{Diaz2022} \Te-based diagnostic. For the strong line metallicity diagnostics, we applied a S/N $>$ 3 cut on all relevant emission lines within each spaxel. For the \Te-based, we used a S/N $>$ 3 cut on the nebular lines, and a S/N $>$ 5 cut on the auroral line. We then further required the resulting metallicity to also have S/N $>$ 3. We note that requiring a measure of the \Te-based metallicity means that only 3.5\% of the spaxels within these regions are selected, and the impacts of this are discussed further in Section \ref{ssubsect:TeSelectionEffects}.

Fig.~\ref{fig:SLvsTe_spaxels} shows that strong-line metallicity estimates are also offset towards lower values compared to \Te-based estimates on sub-\HII region scales, covering a similar region of the parameter space as the stacked \HII region values. 

Overall, these offsets are slightly larger on sub-\HII region scales. For example, the best-fit line to the \citetalias{Pettini2004} O3N2 spaxel measurements suggests an offset of $\sim$0.35 dex below the \Te-based measurements at \logoh\ = 8.8, compared to $\sim$0.26 dex for the \HII stack values. On spaxel scales, the \citetalias{Marino2013} O3N2 diagnostic shows an even flatter distribution, with the O3N2 metallicity almost independent of the \Te-based metallicity. The offsets are increased to $\sim$0.43 dex below the \Te-based values at \logoh\ = 8.8, compared to $\sim$0.31 dex in the \HII region analysis. 

The \DXVI diagnostic again shows a close to linear relationship with the \Te-based method, unlike the other diagnostics, with an offset of $\sim$0.15 dex at \logoh\ = 8.0 increasing to $\sim$0.25 dex at \logoh\ = 8.8.

\begin{figure*}
    \centering
    \includegraphics[width=1\linewidth]{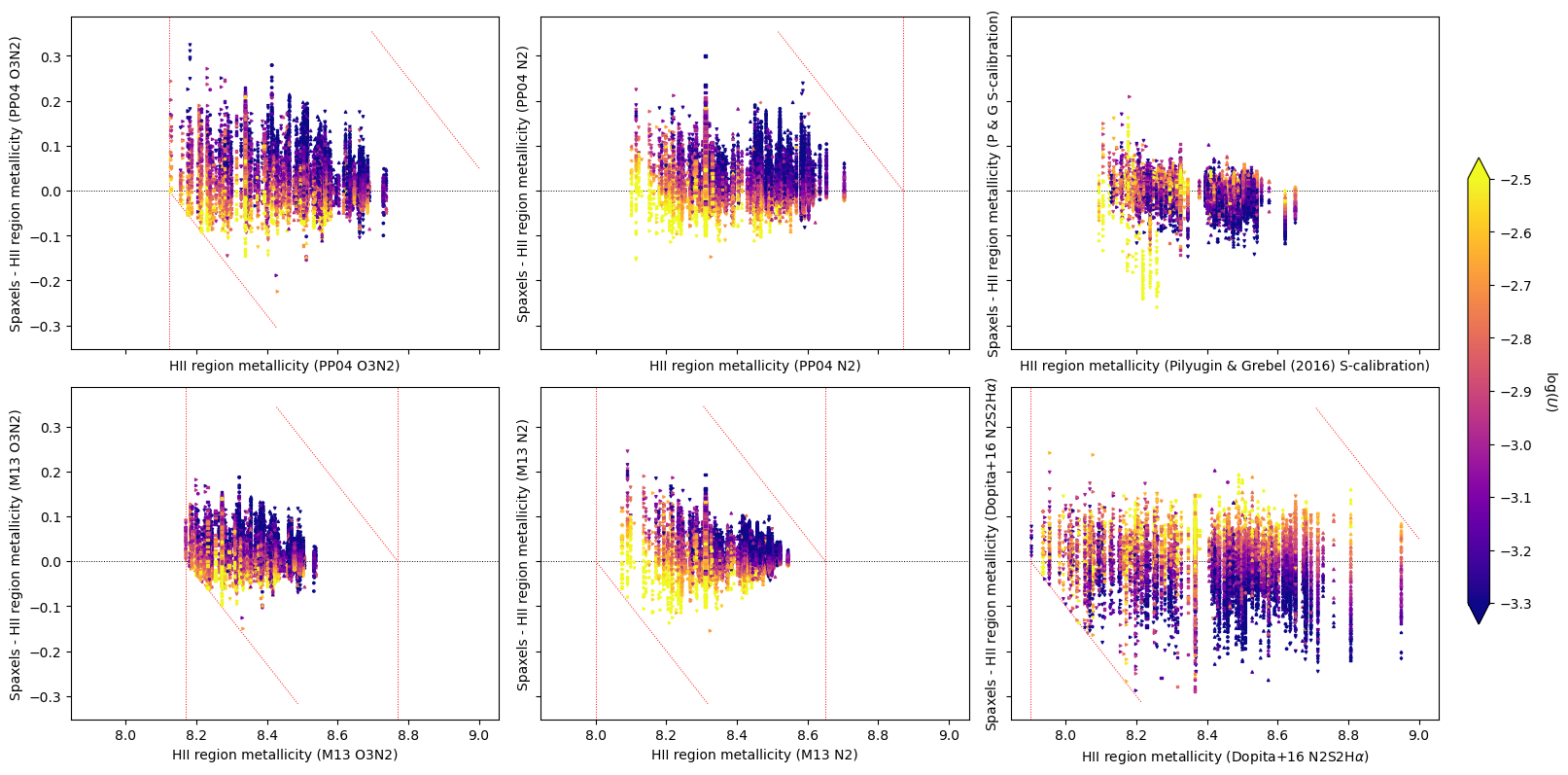}
    \caption{Difference between the metallicity of each spaxel within a given \HII region and the metallicity obtained from the corresponding stacked \HII spectrum for each \HII region in our sample with a \Te-based metallicity. The points are coloured by \logu, as determined using the \protect\cite{Mingozzi2020} diagnostic. Vertical red dotted lines denote the validity limits of the diagnostics, and diagonal red lines show the corresponding limits on the possible differences between the spaxel and region metallicity measurements.}
    \label{fig:diff_spaxels_region}
\end{figure*}

Despite the similarity in the parameter space occupied by single spaxels and stacked \HII regions, of note in Fig.~\ref{fig:SLvsTe_spaxels} is the almost flat distribution of data points for a given galaxy, especially for the O3N2, N2 and S-calibration diagnostics. A possible explanation for this very flat distribution may therefore be that it reflects large variations within galaxies in \logu{} \textit{and} metallicity, but with only a correspondingly small variation in the O3N2 and N2 line ratios. For example, for spaxels in IC5273 with measured \Te-based metallicities, \logu{} covers a range of 0.85~dex and the \Te-based metallicities vary by $\sim 0.7$~dex, whereas the line ratios of log((\oiii/\Hb)/(\nii/\Ha)) cover only 0.4~dex corresponding to just 0.1~dex range in estimated metallicities from both the \citetalias{Pettini2004} and \citetalias{Marino2013} O3N2 diagnostics. When stacking the spaxels within \HII regions, this range in properties is reduced, although there is still evidence of the dependency between O3N2 and N2 line ratios, \logu\ and \Te-based metallicity (Fig. \ref{fig:O3N2lineratio}). This therefore implies that the location of an \HII region in Fig.~\ref{fig:SLvsTe} and Fig.~\ref{fig:SLvsTe_Berg_conversion} will be sensitive to its average \logu{} value, and thus the distribution in \logu{} within the \HII region.

To investigate this further, for each strong line diagnostic considered in Fig.~\ref{fig:SLvsTe}, we plot in Fig.~\ref{fig:diff_spaxels_region} the difference between the stacked and spaxel metallicities ($\Delta{}Z$), against the region's stacked metallicity, colouring each data point by the spaxel \logu. 

Within each region there is a large spread of spaxel metallicities of up to 0.4 dex for the \DXVI diagnostic, reduced to mostly within 0.25 dex for the \citetalias{Marino2013} O3N2 and N2 diagnostics. This range of values is in contrast to works such as \cite{Peimbert2017}, which suggest \HII regions to be chemically homogeneous. There is additionally a strong anti-correlation between $\Delta{}Z$ and \logu{} at fixed stacked metallicity for the O3N2 and N2 diagnostics.

The O3N2 and N2 diagnostics rely on an anti-correlation between the metallicity and \logu, which is reflected in the \logu{} trends seen for these diagnostics in Fig.~\ref{fig:diff_spaxels_region} (left and centre panels, respectively). High values of \logu{} are seen at the low metallicity end on the x-axis in Fig.~\ref{fig:diff_spaxels_region}, and along the y-axis spaxels with lower metallicities have higher \logu{} values than spaxels with higher metallicities, leading to a diagonal colour gradient. 

Conversely, the \DXVI diagnostic should be independent of \logu\ by using a ratio of \nii/\sii, and any observed \logu-dependency with metallicity should thus reflect an intrinsic relation. The bottom right panel in Fig.~\ref{fig:diff_spaxels_region} shows that there is no clear trend with \logu{} at low metallicities, but at higher metallicities (12+log(O/H)$>$8.4) there is a clear positive trend, with spaxels with higher metallicities than that of the overall region also having higher \logu. This suggestion of a positive relationship between metallicity and \logu{} has been previously observed when using the \DXVI \citep{Kruhler2017, Easeman2022}, S-calibration \citep{Kreckel2019, Groves2023}, and the N2O2 (\nii$\lambda$6584 / \oii$\lambda\lambda$3727,29) \citep{Grasha2022} diagnostics.

The \cite{Pilyugin2016} S-calibration shows a much smaller range of metallicities within a single \HII region compared to the \DXVI values, mostly within $\sim$0.2 dex of the measured stack value, with the exception of a few high \logu{} points which lie further away. Although the relationship is less clear, the points seem to show a similar relationship with \logu{} as the \DXVI diagnostic, with no clear relationship at low metallicity, but a positive relationship for metallicities > 8.4.

Aside from the distribution of \logu\ and metallicity within \HII regions in absolute terms, it is also interesting to consider where the spread of data points lie relative to the stacked \HII metallicity (dashed horizontal line). For the O3N2 and N2 diagnostics, the data points lie predominantly above the dashed line ($\sim$70\% of spaxels lie above). The \DXVI and S-calibration instead show the stacked spectra measurements lying closer to the centre of the distribution slightly offset towards higher metallicities than returned from the individual spaxels ($\sim$40\% of spaxels lie above, $\sim$60\% below). Considering the spaxels with metallicity values within 0.01 dex of the stack value, we found the average \logu{} to be fairly consistent across the diagnostics, ranging from -2.97 to -3.00. This implies that for each strong line diagnostic, the \HII region metallicity is weighted towards spaxels with comparable environmental properties.

\begin{figure*}
    \centering
    \includegraphics[width=1\linewidth]{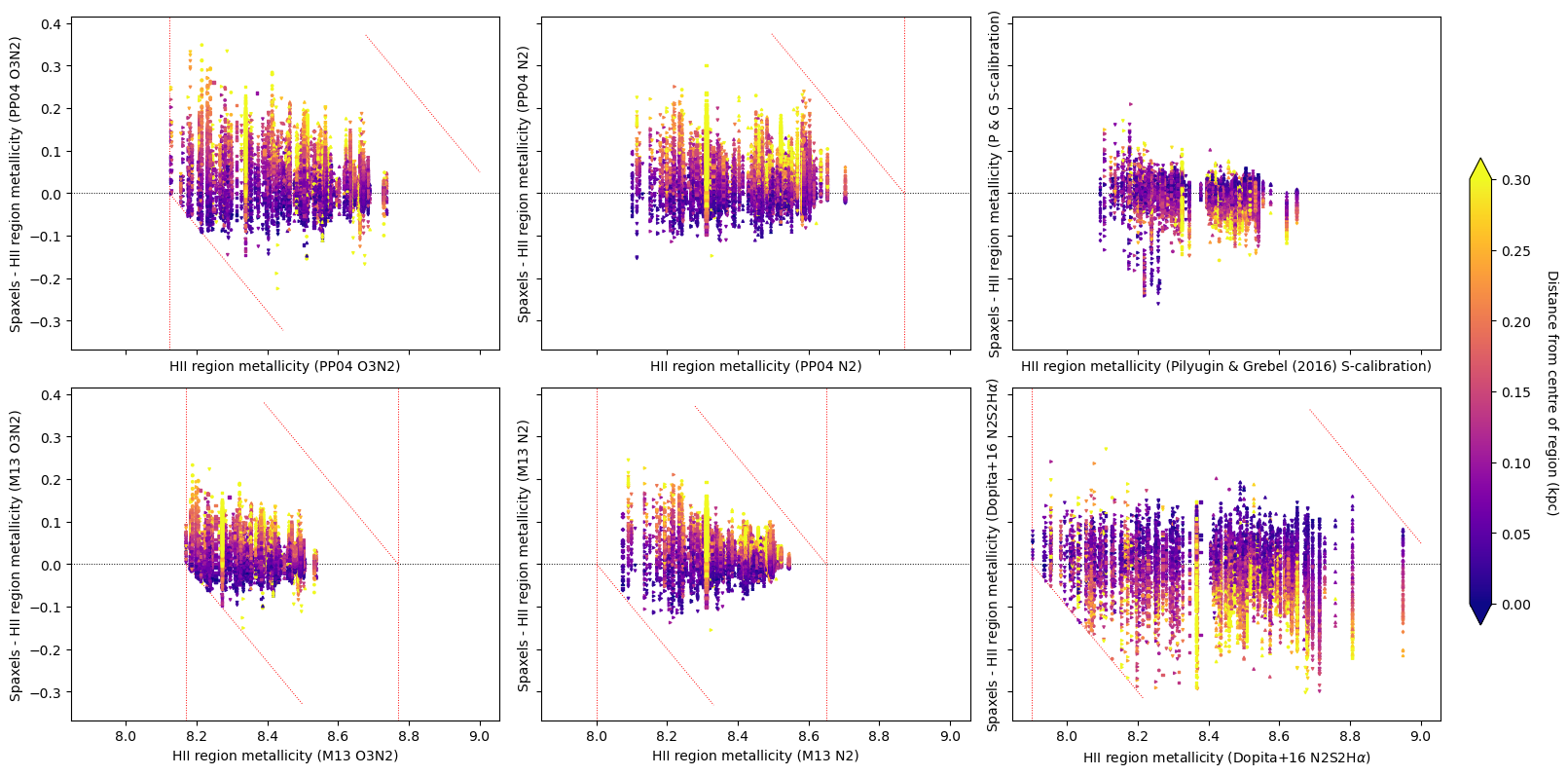}
    \caption{As for Fig.~\ref{fig:diff_spaxels_region}, colouring the points instead by the radial distance of each spaxel from the centre of the region, in kpc.}
    \label{fig:diff_spaxels_region_dist}
\end{figure*}

Given this apparent difference between the diagnostics in how the \HII stacked metallicity compares to the metallicity distribution within the \HII regions, we also investigated the spatial location of the spaxels that carried a higher weight in the stacked \HII metallicities. We did this by colour-coding the points in Fig.~\ref{fig:diff_spaxels_region} by the distance of each spaxel from the centre of the region (Fig.~\ref{fig:diff_spaxels_region_dist}). In doing so, we noticed that the O3N2 and N2 profiles clearly showed profiles inverted to what would be expected, with lower metallicities measured in the centres compared to the outskirts. This was also observed by \cite{Kruhler2017} and \cite{Mao2018}, and similarly to these papers, we find the \DXVI diagnostic to behave as expected, with metallicities increasing towards the centres of \HII regions. The S-calibration and \DXVI diagnostic show negative radial gradients most clearly for regions with metallicity above 8.4. The inverted profiles seen when using the O3N2 and N2 diagnostics is therefore likely a consequence of the positive relationship observed between metallicity and \logu{} in more metal rich \HII regions with the \DXVI and S-calibration diagnostics, contrary to the anti-correlation required by the O3N2 and N2 diagnostics.

For spaxels with metallicities within 0.01 dex of the stack value, the average distance from the centre of the region is also fairly consistent between the diagnostics, ranging from 0.11 to 0.13 kpc. 

As the metallicity of stacked spectra appear to trace the same part of the region and gas with similar levels of excitation for all strong line diagnostics, there is no obvious property on the sub-\HII region scales that can account for the differences observed in Fig. \ref{fig:SLvsTe} between the diagnostics considered. Instead, the distribution in \HII properties echoes the dependencies observed in the stacked spectra between O3N2 and N2 metallicity and \logu.

\subbib

\section{Discussion}
\label{sect:discussion}

Strong line metallicity diagnostics are only indirect tracers of metallicity, and they are generally dependent on multiple environmental properties, most notably \logu\ \citep{Kewley2002}. Diagnostics with either just a weak dependence on \logu\ \cite[e.g. N2O2 at high metallicity;][]{Kewley2002}, or ones which include multiple line ratios that can be used to constrain the various environmental properties \cite[e.g. R$_{23}$;][]{Kobulnicky2004} can therefore be considered more reliable. However, such diagnostics commonly include emission lines at very different wavelength ranges, which are not always accessible with a single spectrograph. In the case of MUSE, the relatively high spectral cut off at the blue end of $\sim 4800$\AA\ makes the important \oii{} line doublet inaccessible for nearby galaxies ($z<0.3$), for which the spatial resolution is highest. Similarly, the \oiiiauroral\ line cannot be detected with MUSE in galaxies at $z<0.1$, meaning that oxygen \Te-based diagnostics cannot be applied. In this paper we have therefore used the \cite{Diaz2022} sulphur \Te-based metallicity diagnostic to investigate the accuracy of several strong line metallicity diagnostics which are applicable to MUSE observations of nearby galaxies ($z<0.03$). This includes the \citet{Dopita2016} \DXVI diagnostic, which was calibrated against the {\sc Mappings} photoionisation code \citep{Dopita2013}, and which has not yet been compared empirically to \Te-based metallicities on a large sample of \HII regions.

In general we find good agreement between the \cite{Diaz2022} sulphur \Te-based metallicities and oxygen \Te-based metallicities \cite{Izotov2006} (see Fig.~\ref{fig:CHAOS_Te_comparison}), but there are significant offsets between the \Te-based metallicities and the strong line diagnostics considered in this paper (Fig.~\ref{fig:SLvsTe}). In this section we consider possible causes for the offsets that we observe, and discuss the implications of our findings.

\subsection{The sulphur-to-oxygen ratio}
\label{subsect:SOratio}

The S/O ratio has been suggested to increase with decreasing metallicity \citep{Dors2016, Diaz2022}, and thus by assuming a constant value, as we do in section~\ref{subsec:ZTe} to convert between 12 + log(S/H) and \logoh\ we could be overestimating \logoh\ at low metallicities (relative to high metallicities). For example, using a combination of star forming galaxies and \HII regions \cite{Dors2016} found the log(S/O) ratio to vary from -1.68 at \logoh\ = 8.0, to -1.90 at \logoh\ = 8.8. Applying such a metallicity-dependent log(S/O) ratio would increase our \Te-based metallicities by $\sim 0.3$~dex at the high metallicity end, further flattening the relation between \Te-based and strong line metallicities, and significantly increasing the offsets we observe between the strong line and \Te-based metallicities in Figs.~\ref{fig:SLvsTe} and \ref{fig:SLvsTe_Berg_conversion}. The log(S/O) values measured for the `DHR' sample presented in \cite{Diaz2022}, which best represents our sample of \HII regions, similarly imply an inverse relation with metallicity, varying from around -1.3 at \logoh\ = 8.0 to around -1.6 at \logoh\ = 8.7. 

In contrast to these results, \cite{Berg2020} found no evidence of a metallicity dependence in log(S/O), instead finding a fairly constant ratio with an average value of log(S/O) = -1.34 for the 190 \HII regions in their sample of four CHAOS galaxies.

With our choice of log(S/O) = -1.57, we find good agreement between the sulphur- and oxygen-based \Te{} diagnostics in Section~\ref{fig:CHAOS_Te_comparison}. If we were instead to apply a varying log(S/O) such as found by \cite{Diaz2022}, discrepancies of $\sim$0.2 dex would be introduced at low metallicities between the \cite{Diaz2022} and \cite{Izotov2006} \Te-based diagnostics, and the differences between \cite{Diaz2022} and \cite{Berg2020} \Te-based metallicities would also increase. If alternatively we used the best-fit constant value of log(S/O) = -1.34 from \cite{Berg2020}, we would reduce the discrepancies between the strong line and sulphur based \Te{} metallicities shown in Fig.~\ref{fig:SLvsTe}, bringing the \DXVI diagnostic in particular in very good agreement. However, unexplained discrepancies of 0.23 dex would then be introduced between the sulphur- and oxygen-based \Te{} diagnostics (Fig.~\ref{fig:CHAOS_Te_comparison}), suggesting that metallicities based on the \cite{Diaz2022} method would no longer provide values in alignment with those returned by oxygen-based methods. We therefore find a fixed value of log(S/O) = -1.57 to be the most appropriate choice.

\subsection{Possible origin of metallicity offsets}
\label{ssubsect:Zoffsets}
Given that the majority of the strong line metallicity diagnostics in Section~\ref{sect:analysis} are calibrated against \Te-based metallicities, it seems surprising that they show such poor agreement with our sulphur \Te-based metallicities. It is therefore reasonable to consider what differences there may be in the range of environmental properties present in the respective calibration samples and in our sample of \HII regions, and how different \Te-based methods compare.

\subsubsection{Selection effects in \Te-based samples}
\label{ssubsect:TeSelectionEffects}

The \oiiiauroral{} line is stronger in high excitation, low metallicity galaxies \cite[e.g.][]{Hoyos2006}, which could bias calibration samples that are based on auroral line detections against low excitation, high metallicity regions. Evidence of such biases can be seen in the \citetalias{Marino2013} \HII sample, where for metallicities above 8.5, there are significantly fewer measurements, and for \citetalias{Pettini2004}, only 3 of their 131 observed \HII regions with \oiiiauroral{} line detections have higher than solar metallicities.

To test whether similar selection effects are present in our \siiiauroral{} line sample, we compared the distribution of \logu\ and strong line metallicities in the full \HII region sample selected in Section~\ref{ssect:hiidentity} by our \hiidentify{} code to the sub-sample of regions with \Te-based measurements. Out of 4408 regions within our sample, 186 (4.2\%) have measured \Te-based metallicities. However we note that our \hiidentify{} code is capable of separating overlapping \HII regions (see Fig.~\ref{fig:Hamapseg}), which would result in a larger number of detected \HII regions, extending down to lower \Ha\ fluxes, compared to other \HII identification codes. For example, from visual inspection of the segmentation maps presented in \cite{Grasha2022} (their fig. 3), where the \texttt{HIIphot} code was used to identify \HII regions, the regions identified appear to be fairly isolated, suggesting that overlapping regions are generally merged together. Evidence of this is present in the galaxy NGC2835, which is present in both our sample and in \cite{Grasha2022}. The MUSE pointing covers just the central $1\arcmin\times 1\arcmin$ part of the galaxy, whereas the data in \cite{Grasha2022} cover an area $\sim 8$ times larger (their fig. 1). Nevertheless, the number of identified \HII regions in \cite{Grasha2022} is only $\sim 30$\% greater. The larger size of our parent sample therefore has the consequence of reducing the fraction of \HII regions with \siiiauroral{} detections than if we had used alternative \HII detection algorithms. 

Similarly to the biases present in \oiiiauroral-based samples, our subset of \siiiauroral-bright \HII regions are indeed shifted towards higher values of \logu{}, covering a range of -3.3 -- -2.2, compared to -4.1 -- -2.2 for the full sample. The metallicities as measured by the strong line diagnostics are also shifted towards lower values, with the maximum metallicities in the sub-sample of \Te-selected regions typically $\sim$0.2 dex below that for the full sample, despite the parent population peaking toward high metallicities.

Similar trends are seen in our spaxel-scale analysis, where in our sample of \HII regions with measured \Te-based metallicities in the stacked spectra, only 3.5\% of spaxels within those regions also have measurements of the \Te-based metallicity. Again the range of \logu{} is reduced, with all spaxels covering a range of -4.0 to -1.8, and those with \Te-based metallicity measurements covering -3.3 to -1.8. The range of metallicity measurements is similarly limited at the high metallicity end, to roughly 0.1 dex below the maximum of that of the parent sample. For example, the \citetalias{Pettini2004} O3N2 values for the parent sample range from 8.15 to 8.8, peaking around 8.6. The sub-sample of regions with \Te-based measurements cover only from 8.15 to 8.65.

In Figs.~\ref{fig:SLvsTe}~--~\ref{fig:SLvsTe_Curti} we saw evidence for strong \logu{} gradients in the O3N2 and N2 comparison plots, whereby \HII regions with lower \logu{} values lay at systematically larger offsets from the line of equality, especially at the higher metallicity end. Given this trend, one could therefore speculate that the large sample of low \logu{} and high metallicity \HII regions omitted by our selection criteria lie closer to the line of equality at the high metallicity end than our \Te-based sample, steepening the slope of the best-fit line. Nevertheless, given that the \oiiiauroral{} \Te-based methods used to calibrate the diagnostics will suffer from similar selection effects, in particular when stacking is not used, as is the case with the \citetalias{Pettini2004} and \citetalias{Marino2013} diagnostics, this bias cannot clearly explain the disagreement that we find in Figs.~\ref{fig:SLvsTe}~--~\ref{fig:SLvsTe_Curti}. Furthermore, the better agreement shown in Fig.~\ref{fig:SLvsTe_Curti} between our \Te-based metallicities and the \cite{Curti2017} O3N2 calibration, which uses stacking to reduce selection effects, implies that our results are not significantly affected by selection effects present in our \siiiauroral-based sample relative to other auroral line-based samples. 

To verify that unaccounted selection effects are not inadvertently causing the flat relation that we find between \siiiauroral\ \Te-based metallicities and the O3N2 and N2 metallicities, we investigated the relation between the sulphur-based and \citetalias{Pettini2004} metallicities in the CHAOS sample, which used fixed slit spectroscopy rather than IFU data. We found the O3N2 and N2 relations to be flatter than our MAD galaxy results shown in Fig.~\ref{fig:SLvsTe}, with respective best-fit slopes of $0.30\pm 0.05$ and $0.31\pm 0.04$ (compared to $0.55\pm 0.05$ and $0.53\pm 0.04$ when using the MAD galaxy sample). The CHAOS sample of \HII regions with \siiiauroral\ detections has predominantly high \logu\ values ($>-3$), implying that CHAOS has a larger \logu\ bias than the MAD sample, and possibly also more biased than the PP04 calibration sample in terms of the range of ionisation parameters covered.

\subsubsection{Differences in electron temperatures}
\citetalias{Marino2013} show their relation between \Te-metallicity and the O3N2 and N2 line ratios next to other relations, including \citetalias{Pettini2004}, in their fig. 2 \& 4, where it is clear that the \citetalias{Marino2013} relation is much shallower. Notably, as illustrated in Fig.~\ref{fig:O3N2lineratio}, at \logoh$\gtrsim$ 8.4 the \cite{Curti2017} O3N2 calibration has a similar slope, but is offset to larger metallicities. The sample of \HII regions used in \citetalias{Marino2013} is larger than used in \citetalias{Pettini2004}, and \citetalias{Marino2013} also use a range of auroral line detections to minimise the impact of temperature-based selection effects. The reason for the systematically lower \citetalias{Marino2013} O3N2 and N2 metallicities at high metallicities, leading to their shallower relation, is not clear. However, it is worth noting that there is scatter in the \citetalias{Marino2013} sample, and in the case of O3N2 (see their fig. 2), the best-fit relation at high metallicities is clearly driven by a tight distribution of data points from \cite{Pilyugin2012}, whereas data from \cite{Perez-Montero2009}, for example, extend to higher metallicities for the same given O3N2 line ratio.

\citetalias{Marino2013} used auroral line detections from various elements to measure the electron temperature of the \HII regions in their sample, using the \cite{Pilyugin2010} method to re-measure \Te-based metallicities. Which of the auroral lines were detected depended largely on the metallicity of the \HII region, with the \nii{} auroral line generally used for metallicities > 8.4, and metallicities at < 8.4 largely measured using the \oiii{} auroral line. When determining metallicities based on a detection of the \nii{} auroral line, their method assumes that \Te$_{, \nii}$ = \Te$_{, \oii}$, and while the two elements do have similar ionisation potentials and can be expected to trace similar parts of the \HII region, various studies have suggested the two temperatures can not be considered to be equal, although there is contention in what the relation should be. \cite{Yates2020} found that especially at higher temperatures, \Te$_{, \nii}$ > \Te$_{, \oii}$ in their observations of \HII regions and galaxies, whereas \cite{Berg2020} found the opposite using the CHAOS sample of \HII regions, suggesting that the relationship between the two temperatures is not well defined, and it is therefore unclear whether it is valid to assume that \Te$_{, \nii}$ = \Te$_{, \oii}$. 
The mixture of auroral lines used in \citetalias{Marino2013} to trace the electron temperature could therefore contribute to the discrepancies we find between the \citetalias{Marino2013} diagnostics and our sulphur \Te-based metallicities, in particular in the case that \Te$_{, \nii}$ > \Te$_{, \oii}$ \citep{Yates2020}, which would imply that \citetalias{Marino2013} have overestimated \Te$_{, \oii}$, and thus underestimated \logoh\ at high metallicities. However, the inhomogeneous sample of \HII regions used in \citetalias{Marino2013} may also contribute to offsets.

\subsubsection{Fixed slit versus IFU measurements}

The \cite{Pilyugin2016} S-calibration has been suggested to return underestimates of the metallicity when used on IFU data, with \cite{Pilyugin2022} finding the results to be underestimated by $\sim$0.06 dex at 12+log(O/H) $\gtrsim$ 8.55, and by $\sim$0.02 below this, on average. They reason that this is due to differences in the intensities of each line in single-slit and IFU observations, due to the limitations of single-slit observations meaning that the entire \HII region is not always captured in a single observation. They suggest this introduces systematic effects when their diagnostic, which was calibrated using single-slit observations, is used on IFU data. This could be expected to therefore also affect the \citetalias{Pettini2004} diagnostics, which were calibrated against single slit data, and the \citetalias{Marino2013} diagnostics, which were calibrated using a combination of IFU data and single slit observations. However, this can only account for some of the offsets observed between the diagnostics in our analysis, as we observed systematic offsets of up to 0.3 dex at high metallicities. Furthermore, our analysis on the spaxel-level data indicate that the \HII stacked spectra are predominantly weighted by the central regions of the \HII region, implying that emission missed in single slit spectra from the outer regions of the \HII region are unlikely to contribute greatly to the measured line fluxes.

\subsection{Large and small scale metallicity gradients}

The strong line O3N2, N2 and S-calibration diagnostics underestimate the metallicity at high metallicities by $\sim$0.3 dex, which could have implications on measurements of the metallicity gradient, for example imposing a flattened inner profile for a galaxy with a negative metallicity gradient. The \DXVI diagnostic shows no evidence of a metallicity-dependent offset from \Te-based measurements, and no clear \logu{} dependence in the discrepancies between the strong line and \Te-based measurements. This could explain why, for a sample of galaxies observed in the MaNGA survey \citep[Mapping Nearby Galaxies at Apache Point Observatory;][]{Bundy2015}, \cite{Yates2021} found evidence of flattened inner profiles in more massive galaxies when O3N2-based diagnostics were used, which were not seen in results from the \DXVI diagnostic, as well as finding that the N2 diagnostic also returned much flatter profiles than the \DXVI diagnostic.

On sub-\HII region scales, we find in Figs. \ref{fig:diff_spaxels_region} and \ref{fig:diff_spaxels_region_dist} that the O3N2 and N2 diagnostics show clear inverted metallicity profiles, with metallicity appearing to decrease towards the centre of the \HII regions. The \DXVI diagnostic, however, shows decreasing metallicities from the centres of the regions out to the outskirts, which is in line with expectations. Exploring the relation between metallicity and \logu, we find that at high metallicities, the \DXVI diagnostic shows a clear positive relationship between \logu{} and metallicity, which would explain the inverted profiles shown by the O3N2 and N2 diagnostics, as they rely on an anti-correlation between metallicity and \logu{} observed in \HII region and galaxy samples.

\subsection{Optimal metallicity diagnostics with MUSE}
On both \HII region and sub-\HII region scales, the \DXVI diagnostic results are offset from the \Te-based values, however this discrepancy has no apparent dependence on either metallicity or ionisation parameter, unlike the other diagnostics tested. We therefore find that this diagnostic produces the most reliable results when used on spatially resolved MUSE observations of nearby galaxies.

We note, however, that the \HII regions used in our analysis typically account for just $\sim$10 \% of the overall \Ha\ luminosity of the galaxy within the MUSE field of view, therefore these results may not be descriptive of trends seen on whole-galaxy scales.

Further study of the diagnostics will be possible with the proposed BlueMUSE instrument \citep{BlueMUSE}. BlueMUSE will have a spectral range covering much shorter wavelengths (3500 - 6000 \AA), and will provide high spatial resolution IFU data with coverage of the \oii$\lambda\lambda$3727,29 and \oiiiauroral\ lines, allowing for \Te-based diagnostics reliant on the oxygen lines to be used, removing the need for any conversion between sulphur and oxygen abundances, and allowing a larger range of strong line diagnostics to be investigated. 

\subbib

\section{Conclusions}
\label{sect:conclusion}

We have used a sample of 671 \HII regions from 36 galaxies observed as part of the MUSE Atlas of Disks survey, to assess the optimal strong line diagnostic for use with MUSE data. We additionally release a catalogue of 4408 \HII regions identified within this sample using our newly-developed python tool \hiidentify\footnote{\url{https://hiidentify.readthedocs.io/en/latest/}}, with segmentation maps and tables of emission line strengths made available (see Appendix \ref{appendix:catalogue}). By comparing the results from strong line diagnostics to \Te-based measurements, we find:

\begin{itemize}
\item The \citetalias{Pettini2004} O3N2, \citetalias{Pettini2004} N2, and \cite{Pilyugin2016} S-calibration diagnostics all show consistent results, with a sub-linear relationship leading to agreement around \logoh\ = 8.2, but an offset of $\sim$ 0.3 dex below the \Te-based values at \logoh\ = 8.8. The O3N2 and N2 diagnostics additionally show strong \logu{} dependence in the offsets from the \Te-based values. The \DXVI diagnostic shows the greatest level of agreement with the \Te-based values, with the slope of the best-fit line being consistent with a 1:1 relationship. However, this diagnostic has a systematic offset of $\sim$0.2 dex, which, unlike the O3N2 and N2 diagnostics, has no clear dependence on \logu{} or metallicity. 

 \item The O3N2 and N2 diagnostics presented by \cite{Marino2013} show the largest differences with the \Te-based measurements at high metallicities, with these strong line diagnostics returning values over a significantly reduced range compared to the \Te-based values, due to the much steeper relationship between the line ratios and metallicity (see Fig.~\ref{fig:O3N2lineratio}).

 \item This comparison on \HII region scales suggests that when measuring the metallicity of \HII regions for galaxies observed with MUSE, the \DXVI diagnostic provides the most accurate measurement of the metallicity. 

 \item We used the CHAOS sample \citep{Berg2020} to assess the validity of using the sulphur \Te-based method presented by \cite{Diaz2022} to measure \logoh, finding good agreement with the \cite{Izotov2006} oxygen \Te-based method. However, when comparing the results to the published results from applying the multi-zone \Te-based method presented in \cite{Berg2020}, there is a significant offset at lower metallicities (-0.37 dex at \logoh\ = 8.0). If we use our fitted relationship to `convert' our measured \Te-based metallicities for the MAD data to predicted values for if the \cite{Berg2020} method could be used on our data, the resulting comparisons between the strong line and \Te-based diagnostics are significantly changed, with reduced metallicity-dependence in the offsets for the O3N2, N2 and S-calibration diagnostics, but increased dependence for the N2S2H$\alpha$.

 \item On sub-\HII region scales, we find clear evidence of inverted metallicity profiles when using the O3N2 and N2 diagnostics, which are not seen when the S-calibration or \DXVI diagnostic is used. For the \DXVI diagnostic, we find a clear positive relationship between \logu{} and metallicity at high metallicities, in contrast to the negative relationship required by the O3N2 and N2 diagnostics. This could therefore explain the inverted profiles observed within \HII regions, as has been suggested by \cite{Kruhler2017} and \cite{Mao2018}.
 
 \item Stacked spectra from \HII regions appear to generally be weighted towards the inner part of \HII regions, where the \logu{} is highest. The differing relationships between \logu{} and metallicity between the diagnostics may therefore suggest a reason for the disagreement between the N2S2H$\alpha$, and O3N2 and N2 diagnostics. 
 \end{itemize}

\subbib

\section*{Acknowledgements}

The authors would like to thank {\'{A}}ngeles D{\'{i}}az for an interesting and enlightening conversation surrounding the use of sulphur \Te-based metallicity diagnostics, and the S/O ratio. We also thank Dr Julian Stirling for his guidance on publishing \hiidentify{}, and the referee for their feedback.

\section*{Data Availability}

Data from the MUSE Atlas of Disks (MAD) was used, and can be accessed from the ESO archive science portal\footnote{\url{http://archive.eso.org/scienceportal/home}}. The \HII region catalogue created and analysed as part of this work is provided as an online table (see Appendix \ref{appendix:catalogue}), with the segmentation maps from \hiidentify{} available at \url{https://doi.org/10.6084/m9.figshare.22041263}.



\bibliography{PhD,Additional_bibliography} 



\appendix

\section{Inverted metallicity profiles observed with O3N2 and N2 diagnostics}
\label{appendix:invertedprofiles}

\begin{figure}
    \centering
    \includegraphics[width=1\linewidth]{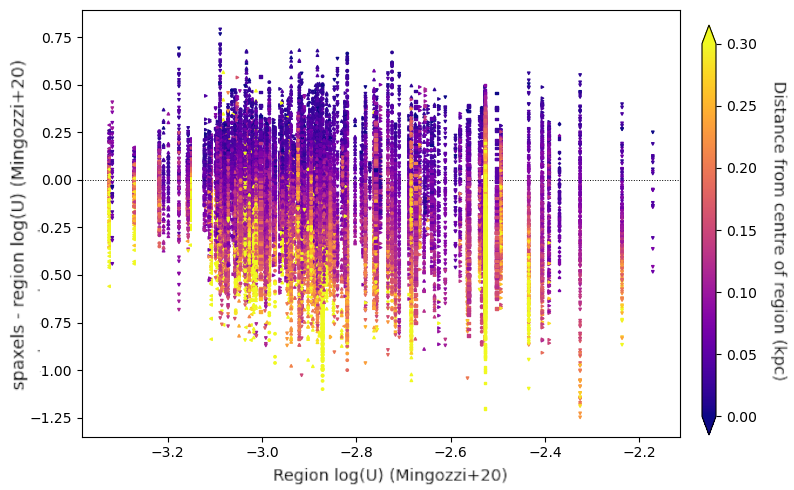}
    \caption{As for Fig.~\ref{fig:diff_spaxels_region}, comparing the \logu{} measured in each spaxel within a region, to the values returned from the region's stacked spectrum, coloured by the distance to the centre of the region.}
    \label{fig:diff_spaxels_region_logU}
\end{figure}

Repeating the analysis described in Section \ref{subsect:spaxels}, but instead plotting the \logu{} that we measured, we see in Fig.~\ref{fig:diff_spaxels_region_logU} that the stacked \HII region value is also weighted towards the value of spaxels near the centre of the region. For spaxels with \logu{} values within 0.01 of that of the region stack, the average distance from the centre of the region is 0.11 kpc, similar to the location of spaxels that contribute most to the stacked metallicities (Fig. \ref{fig:diff_spaxels_region_dist}). 30\% of the spaxels have \logu{} values higher than that of the stacked region value, 70\% have lower values, suggesting the stacked value is slightly shifted towards higher \logu{} than the distribution of values within the region. Smooth gradients in \logu{} can be seen for all regions, with high values of \logu{} found in the centres of regions, decreasing with radius. 

\section{MAD catalogue}
\label{appendix:MAD}

Table \ref{tab:MADprops} details the global properties of the MAD galaxies used in this work, with information compiled from \cite{Erroz-Ferrer2019} and \cite{Salo2015}.

\begin{table*}
\caption{Sample information, compiled from \protect\cite{Erroz-Ferrer2019} and \protect\cite{Salo2015}. }
\centering
\begin{tabular}{lrrrrrrrrr} 
\multicolumn{1}{c}{Name} & \multicolumn{1}{c}{z} & \multicolumn{1}{c}{D (Mpc)} & \multicolumn{1}{c}{$R\sub{e}$ (arcsec)} & \multicolumn{1}{c}{log(M$_*$/M$_\odot$)} & \multicolumn{1}{c}{SFR (M$_\odot$/yr)} & \multicolumn{1}{c}{PA (deg)} & \multicolumn{1}{c}{q} & \multicolumn{1}{c}{RA (deg)} & \multicolumn{1}{c}{DE (deg)} \\
\hline
NGC4030    & 0.004887 & 29.9 & 31.8 & 11.18 & 11.08 & 29.6  & 0.805 & 180.098510 & -1.099960  \\
NGC3256    & 0.009354 & 38.4 & 26.6 & 11.14 & 3.10  & 60.0  & 0.680 & 156.963624 & -43.903748 \\
NGC4603    & 0.008647 & 32.8 & 44.7 & 11.10 & 0.65  & 40.0  & 0.580 & 190.230042 & -40.976389 \\
NGC3393    & 0.012509 & 55.2 & 21.1 & 11.09 & 7.06  & -20.0 & 0.720 & 162.097750 & -25.162056 \\
NGC1097    & 0.004240 & 16.0 & 55.1 & 11.07 & 4.66  & 126.5 & 0.567 & 41.579410  & -30.274910 \\
NGC289     & 0.005434 & 24.8 & 27.0 & 11.00 & 3.58  & 123.7 & 0.658 & 13.176520  & -31.205830 \\
IC2560     & 0.009757 & 32.2 & 37.8 & 10.89 & 3.76  & 40.0  & 0.380 & 154.077985 & -33.563795 \\
NGC5643    & 0.003999 & 17.4 & 60.7 & 10.84 & 1.46  & 87.5  & 0.670 & 218.169765 & -44.174406 \\
NGC3081    & 0.007976 & 33.4 & 18.9 & 10.83 & 1.47  & 70.0  & 0.480 & 149.873080 & -22.826277 \\
NGC4941    & 0.003696 & 15.2 & 64.7 & 10.80 & 3.01  & 23.5  &       & 196.054760 & -5.551620  \\
NGC5806    & 0.004533 & 26.8 & 27.2 & 10.70 & 3.61  & 166.6 & 0.487 & 225.001800 & 1.891270   \\
NGC3783    & 0.009730 & 40.0 & 27.7 & 10.61 & 6.93  & -15.0 & 0.620 & 174.757342 & -37.738670 \\
NGC5334    & 0.004623 & 32.2 & 51.2 & 10.55 & 2.45  & 15.1  & 0.735 & 208.226960 & -1.114760  \\
NGC7162    & 0.007720 & 38.5 & 18.0 & 10.42 & 1.73  & 12.7  & 0.426 & 329.912720 & -43.306220 \\
NGC1084    & 0.004693 & 20.9 & 23.8 & 10.40 & 3.69  & 38.4  & 0.606 & 41.499690  & -7.578640  \\
NGC1309    & 0.007125 & 31.2 & 20.3 & 10.37 & 2.41  & 78.9  & 0.891 & 50.527320  & -15.400070 \\
NGC5584    & 0.005464 & 22.5 & 63.5 & 10.34 & 1.29  & 156.9 & 0.662 & 215.599210 & -0.387450  \\
NGC4900    & 0.003201 & 21.6 & 35.4 & 10.24 & 1.00  & 140.8 & 0.834 & 195.163120 & 2.501460   \\
NGC7496    & 0.005365 & 11.9 & 66.6 & 10.19 & 1.80  & 33.2  &       & 347.447050 & -43.428020 \\
NGC7552    & 0.005500 & 14.8 & 26.0 & 10.19 & 0.59  & 38.5  & 0.586 & 349.044860 & -42.584830 \\
NGC1512    & 0.002995 & 12.0 & 63.3 & 10.18 & 1.67  & 74.5  & 0.591 & 60.976170  & -43.348850 \\
NGC7421    & 0.005979 & 25.4 & 29.6 & 10.09 & 2.03  & 64.6  & 0.739 & 344.226320 & -37.347200 \\
ESO498-G5  & 0.008049 & 32.8 & 19.8 & 10.02 & 0.56  & -25.0 & 0.740 & 141.169458 & -25.092694 \\
NGC1042    & 0.004573 & 15.0 & 63.7 & 9.83  & 2.41  & 6.7   & 0.779 & 40.099900  & -8.433650  \\
IC5273     & 0.004312 & 15.6 & 33.8 & 9.82  & 0.83  & 49.0  & 0.638 & 344.861330 & -37.702880 \\
NGC1483    & 0.003833 & 24.4 & 19.0 & 9.81  & 0.43  & 125.7 & 0.735 & 58.198370  & -47.477520 \\
NGC2835    & 0.002955 & 8.8  & 57.4 & 9.80  & 0.38  & -22.0 & 0.870 & 139.470458 & -22.354667 \\
PGC3853    & 0.003652 & 11.3 & 73.1 & 9.78  & 0.35  & 106.4 & 0.853 & 16.270320  & -6.212380  \\
NGC337     & 0.005490 & 18.9 & 24.6 & 9.77  & 0.57  & 119.6 & 0.633 & 14.958710  & -7.577960  \\
NGC4592    & 0.003566 & 11.7 & 37.9 & 9.68  & 0.31  & 94.6  & 0.298 & 189.828250 & -0.531840  \\
NGC4790    & 0.004483 & 16.9 & 17.7 & 9.60  & 0.39  & 86.2  & 0.646 & 193.716380 & -10.247820 \\
NGC3513    & 0.003983 & 7.8  & 55.4 & 9.37  & 0.21  & 67.6  & 0.722 & 165.942000 & -23.245480 \\
NGC2104    & 0.003873 & 18.0 & 16.5 & 9.21  & 0.24  & 167.3 & 0.547 & 86.769880  & -51.552950 \\
NGC4980    & 0.004767 & 16.8 & 13.0 & 9.00  & 0.18  & 169.6 & 0.516 & 197.292010 & -28.641790 \\
NGC4517A   & 0.005087 & 8.7  & 46.8 & 8.50  & 0.10  & 23.8  & 0.573 & 188.117170 & 0.389750   \\
ESO499-G37 & 0.003186 & 18.3 & 18.3 & 8.47  & 0.14  & 40.0  & 0.640 & 150.924333 & -27.027806
\label{tab:MADprops}
\end{tabular}
\end{table*}

\section{\HII region catalogue produced using \hiidentify{} for the MAD galaxies}
\label{appendix:catalogue}

Using \hiidentify, the python tool we developed to identify \HII regions within galaxies\footnote{Available at: \url{https://hiidentify.readthedocs.io/en/latest/}}, we identified a total of 4408 \HII regions within 36 galaxies from the MAD sample. For each of these regions, we stacked all spectra within the regions, and fitted the emission lines, as described in Section \ref{subsect:fitting}. We provide fluxes for the \Hb, \oiii$\lambda\lambda$4959,5007, \oi$\lambda\lambda$6302,65, \siiiauroral, \nii$\lambda\lambda$6548,84, \Ha, \sii$\lambda\lambda$6717,31 and \siii$\lambda$9070 lines in an online table, an excerpt of which can be seen in Table \ref{table:fluxes}. The segmentation maps denoting which pixels belong to each region can be found at \url{https://doi.org/10.6084/m9.figshare.22041263}. These segmentation maps have 4 extensions, with the first providing a region ID for each pixel, the second a measure of the distance of the pixel from the peak of the region in kpc, the third a map of the peak positions, and the fourth shows the \Ha{} flux within the identified regions.

\begin{table*}
\caption{Example dust-corrected fluxes in 10$^{-20}$ erg/s/cm$^2$ measured for the 4408 \HII regions identified within our sample using \hiidentify.}
\centering
\begin{tabular}{ccrrrrrrrr}
Galaxy name & Region ID & \multicolumn{1}{c}{\Hb} & \multicolumn{1}{c}{\oiii$\lambda$4959} & \multicolumn{1}{c}{\oiii$\lambda$5007} & \multicolumn{1}{c}{\siiiauroral} & \multicolumn{1}{c}{\nii$\lambda$6548} & \multicolumn{1}{c}{\Ha} & \multicolumn{1}{c}{\nii$\lambda$6584} & ...\\
\hline
NGC4030 & 1 & 580963.98 & 24145.05 & 71701.19 & -- & 218013.01 & 2145082.91 & 670310.83 & ... \\
NGC4030 & 2 & 422113.73 & 13597.15 & 40378.11 & -- & 163837.02 & 1489530.26 & 503739.32 & ...\\
NGC4030 & 3 & 174188.67 & 7093.77 & 21065.66 & 1296.39 & 61948.63 & 593299.01 & 190469.54 & ... \\
NGC4030 & 4 & 612040.38 & 17424.16 & 51742.82 & -- & 188234.22 & 2266623.49 & 578751.85 & ...\\
NGC4030 & 5 & 646705.79 & 15994.38 & 47496.94 & 2313.41 & 227074.56 & 2291716.1 & 698171.78 & ...\\

\label{table:fluxes}
\end{tabular}
\end{table*}

\begin{table*}
\caption{Example uncertainties on the measured fluxes for the 4408 \HII regions identified within our sample using \hiidentify. }
\centering
\begin{tabular}{ccrrrrrrrr}
Galaxy name & Region ID & \multicolumn{1}{c}{\Hb} & \multicolumn{1}{c}{\oiii$\lambda$4959} & \multicolumn{1}{c}{\oiii$\lambda$5007} & \multicolumn{1}{c}{\siiiauroral} & \multicolumn{1}{c}{\nii$\lambda$6548} & \multicolumn{1}{c}{\Ha} & \multicolumn{1}{c}{\nii$\lambda$6584}  & ...\\
\hline
NGC4030 & 1 & 60880.78 & 6458.79 & 5789.58 & -- & 30554.0 & 115493.37 & 25870.24 & ... \\
NGC4030 & 2 & 35287.18 & 3236.24 & 2931.34 & -- & 16787.9 & 58892.73 & 14378.95 & ...  \\
NGC4030 & 3 & 22356.62 & 2457.38 & 2191.55 & 1046.3 & 9251.78 & 34267.21 & 7797.34 & ...  \\
NGC4030 & 4 & 31021.66 & 3103.83 & 3058.51 & -- & 19416.09 & 86717.73 & 16463.99 & ...  \\
NGC4030 & 5 & 48424.97 & 3795.56 & 3499.44 & 1523.36 & 27873.0 & 108669.28 & 23925.99 & ... \\
\end{tabular}
\end{table*}

\section{Comparing the results from \hiidentify{} to \hiiphot{}}
\label{app:hiidentifyhiiphot}

\begin{figure}
    \centering
    \includegraphics[width=1\linewidth]{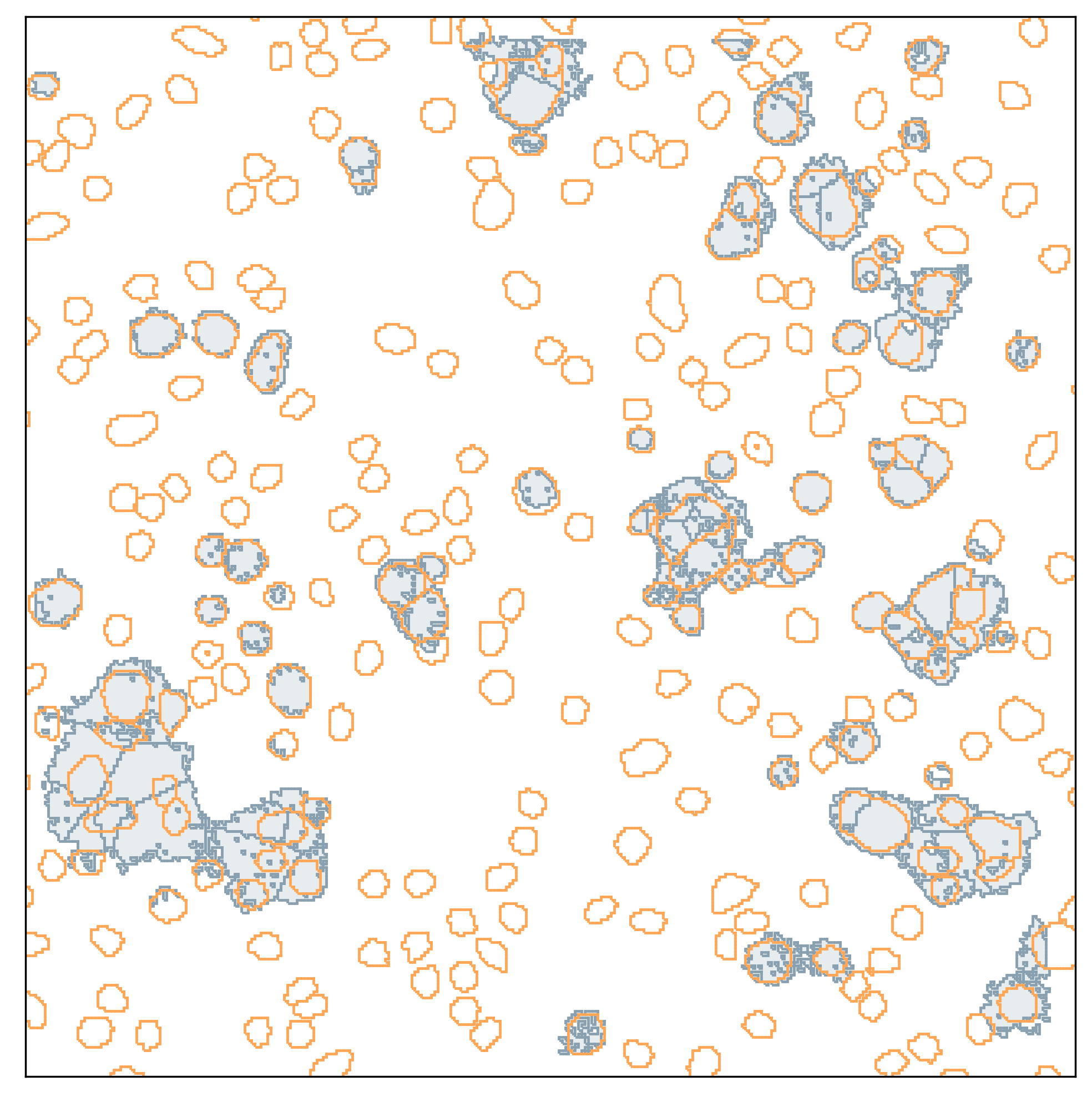}
    \caption{Field of view of the MAD observation of the centre of NGC2835, showing the regions identified with \hiidentify{} (grey outlines and shaded) and with \hiiphot{} (orange outlines) as provided by \protect\cite{Groves2023}.}
    \label{fig:overlaid_regions}
\end{figure}

A number of codes exist to identify \HII regions within galaxies, and following the release of the MUSE-PHANGS \citep[Physics at High Angular resolution in Nearby GalaxieS;][]{Emsellem2022} catalogue by \cite{Groves2023}, we compare the results from our code to those from \hiiphot{} for NGC2835, which has been analysed in both surveys. 

The methodologies of the two codes differ slightly, for example \hiiphot{} begins with an initial guess of the shape of the region, chosen from a set of 6 possible models, and grows the regions iteratively based on a slowly declining flux threshold. \hiidentify{} proceeds by identifying just the brightest pixel within each region, and growing outwards from that point. As the codes iterate outward, different conditions are used to terminate the growth of the region. For the results from using \hiiphot{} presented in \cite{Groves2023}, the edges of the regions were set using the gradient of the \Ha{} surface brightness, with a fixed value
used for all galaxies. For our analysis using \hiidentify{}, we terminate the growth of a region once the flux drops to a defined background value, determined individually for each galaxy. 

To compare the results of the two codes, we matched the area observed in our MAD data to the wider field of view of the PHANGS observations, achieved by mosaicking 7 MUSE pointings. As shown in  Fig.~\ref{fig:overlaid_regions} the \hiiphot{} code (orange outlines) appears to return very isolated regions compared to \hiidentify{} (grey regions), suggesting that neighboring regions may be merged together by \hiiphot{}. The \hiiphot{} code also identifies a greater number of regions, including a large number within areas removed in our analysis due to having \haew{} $<$ 6 \AA. Some of the regions appear to match up well between the results from the two codes, but for other regions it appears that \hiidentify{} has split what was considered as one region by \hiiphot{} into multiple regions. Some of the regions identified by \hiidentify{} are also larger, which may be due to the difference in how the growth of the regions are terminated in the two identification codes, with \hiidentify{} capturing more of the diffuse emission around the edges of the region.

\begin{figure}
    \centering
    \includegraphics[width=1\linewidth]{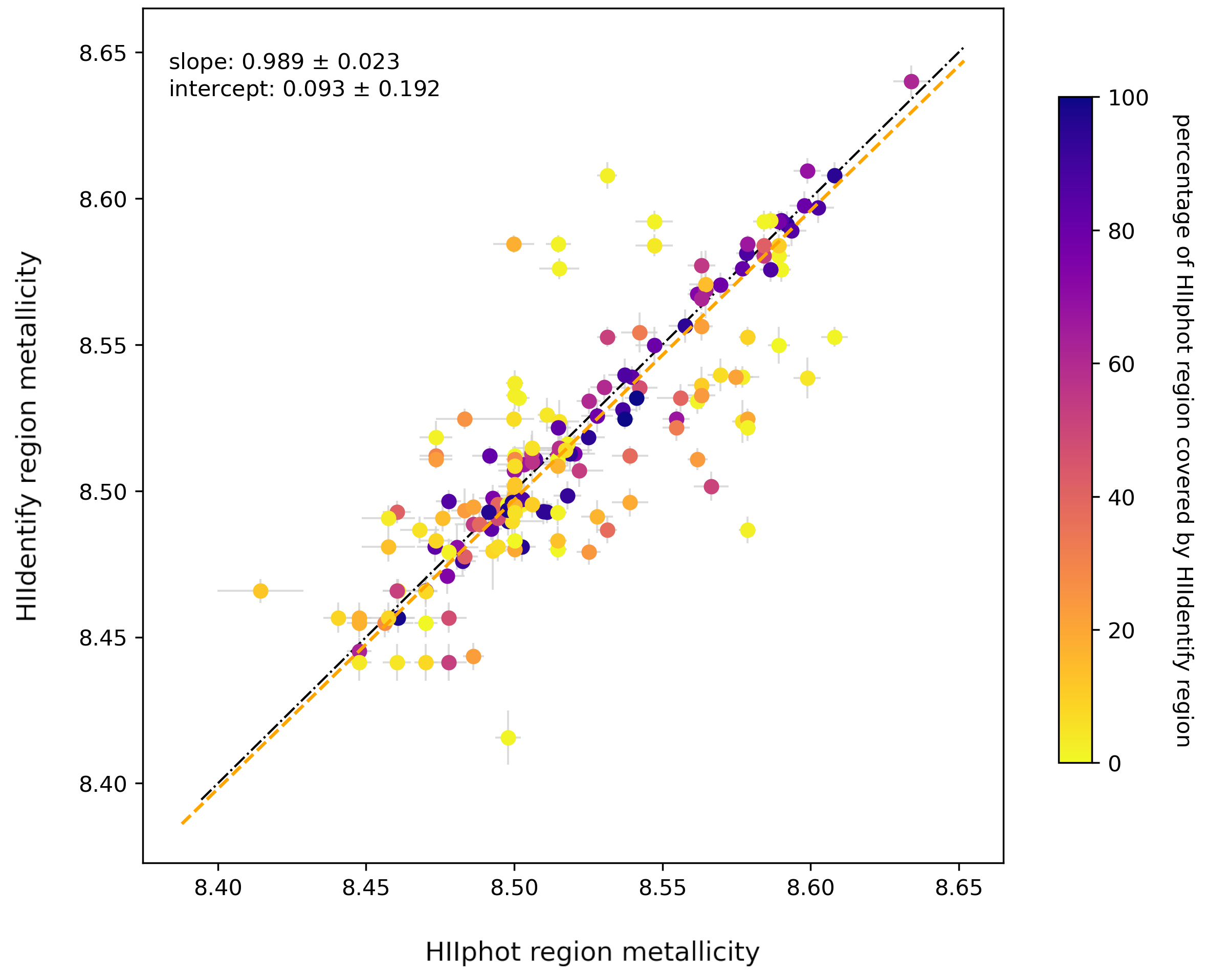}
    \caption{For each region identified by \hiiphot{} which overlaps with at least one \hiidentify{} region, we plot the S-calibration metallicity for the \hiiphot{} region on the x-axis, and the metallicity for each overlapping region as identified by \hiidentify{} on the y-axis. The points are coloured by the percentage of the \hiiphot{} region covered by the \hiidentify{} region, with the orange line of best-fit to the data also weighted by this.}
    \label{fig:Scal_hiiphot_hiidentify}
\end{figure}

Due to the differences in the identification of regions between the two codes, we then explored whether this led to any differences in the measured metallicity. Using our MAD data and the \hiiphot{} segmentation map, we stacked and fitted the spectra within regions identified by \hiiphot{} following the same process as described in Section~\ref{subsect:fitting}. Then for each \hiiphot{} region, we selected every region identified by \hiidentify{} with any overlapping pixels. To compare the metallicities returned, in Fig.~\ref{fig:Scal_hiiphot_hiidentify} we plot the S-calibration metallicity from the \hiiphot{} regions on the x-axis, against the metallicity of each overlapping \hiidentify{} region on the y-axis. The dot-dashed black line shows the 1:1 relation between the two and the points are coloured by the percentage of the \hiiphot{} region covered by the \hiidentify{} region, which is also used to weight the best-fit to the data shown as the orange dashed line. 

Fig.~\ref{fig:Scal_hiiphot_hiidentify} shows very good agreement between the S-calibration metallicities returned for the regions identified by \hiiphot{} and the overlapping \hiidentify{} regions, especially for the regions which share a large fraction of the region's pixels (purple points). We find a similar amount of consistency for all diagnostics used in our analysis. This implies that the nebular line fluxes that we measure are consistent with the published MUSE-PHANGS measurements, and that our results are robust and show little dependence on the code used to identify the \HII regions.

\subbib



\bsp	
\label{lastpage}
\end{document}